\pdfoutput=1

\documentclass[12pt]{article}

\usepackage{latexsym}
\usepackage{amsmath}
\usepackage{amsfonts}
\usepackage{amssymb}
\usepackage{psfrag}
\usepackage{graphicx}
\usepackage{graphics}
\usepackage{amsthm}
\usepackage{mathtools}
\usepackage{enumitem}
\usepackage{geometry}
\usepackage{cite}
\usepackage{hyperref}

\hypersetup{
    colorlinks=true,
    linkcolor=blue,
    filecolor = blue
, urlcolor = blue,
citecolor = blue
}

\newtheorem{theorem}{Theorem}
\newtheorem{lemma}[theorem]{Lemma}

\renewcommand{\eqref}[1]{(\ref{#1})}

\newcommand{\blt}[1]{\mathbf #1}
\newcommand{\bs}[1]{\boldsymbol #1}

\newtheorem{manualtheoreminner}{Lemma}
\newenvironment{manualtheorem}[1]{%
  \renewcommand\themanualtheoreminner{#1}%
  \manualtheoreminner
}{\endmanualtheoreminner}

\begin{document}

\title{Tests of Linear Hypotheses using 
Indirect Information}

\title{Tests of Linear Hypotheses using 
Indirect Information} 
\author{Andrew McCormack and Peter Hoff \\
Department of Statistical Science \\
Duke University} 
\date{\today}

\maketitle

\begin{abstract}
In multigroup data settings with small within-group 
sample sizes, standard $F$-tests of group-specific linear hypotheses 
can have low power, particularly 
if the within-group sample sizes are not large relative to the 
number of explanatory variables. 
To remedy this situation, in this article we
derive alternative test statistics based on information-sharing 
across groups. Each group-specific test has 
 potentially much larger power than the standard $F$-test, 
while still exactly maintaining a target type I error rate if 
the hypothesis for the group is true. The proposed test 
for a given group
uses a statistic that has optimal 
marginal power under a prior distribution derived 
from the data of the other groups. This statistic approaches 
the usual $F$-statistic as the prior distribution becomes more diffuse, 
but approaches a limiting ``cone'' test statistic as the prior 
distribution becomes extremely concentrated. We compare the power and $p$-values of the cone test to that of the $F$-test in some high-dimensional asymptotic scenarios. An analysis of educational outcome data is provided, demonstrating empirically that the proposed test is more powerful than the $F$-test. 

\smallskip
\noindent {\bf Keywords:}  empirical Bayes, $F$-test, frequentist testing, 
hierarchical model,
invariant test, multilevel data, small area estimation. 
\end{abstract}

\section{Introduction}

\label{sec:intro}   

Multigroup data analysis often occurs 
through the lens of 
separate but related linear regression models for each 
of several groups.
 For example, letting $y_{i,j}$ be a real-valued 
outcome and $\blt x_{i,j}\in \mathbb R^p$ be a vector of features 
for the $i$th subject in group $j$, the relationship between 
$y_{i,j}$ and $\blt x_{i,j}$ is often explored 
via the assumption that 
$ y_{i,j} = \blt x_{i,j}^\top \bs \beta_j + \sigma_j \epsilon_{i,j}, $
where the $\epsilon_{i,j}$'s are standard normal random variables, 
independent within and across groups. 
Letting $\blt y_j\in \mathbb R^{n_j}$ be the  vector of outcomes
in group $j$, and $\blt X_j\in \mathbb R^{n_j\times p}$ 
be the matrix of explanatory variables, 
this model can be expressed as 
\begin{equation}
\label{eqn:wgm}
\blt y_j \sim N_{n_j} (\blt X_j \bs \beta_j , \sigma^2_j \blt I),  
\end{equation}
independently across groups $j=1,\ldots, m$.

Estimators of $\bs\beta_1 \ldots, \bs\beta_p$ can be broadly categorized 
as being either ``direct'' or ``indirect''. A direct estimator 
of ${\bs\beta}_j$ is one that makes use of data only from group $j$, 
such as the ordinary least squares (OLS) estimator
$\hat{\bs\beta}_j =  ( \blt X_j^\top \blt  X_j)^{-1} \blt X_j^\top \blt y_j$. 
While $\hat{\bs\beta}_j$ has minimum variance among unbiased estimators, 
if $n_j$ is small then it may be preferable to reduce variance further
by introducing bias in the form of 
``indirect'' information from the other groups. 
Such estimators are often derived by imagining 
a normal model for across-group 
variation, 
\begin{equation}
\bs \beta_1,\ldots, \bs\beta_m \sim \text{i.i.d.} \  N_p(\bs \beta_0 , \Psi).
\label{eqn:agm}
\end{equation} 
This across-group model is sometimes referred to as a
``linking model'' in the small area estimation literature, as it
``links'' the group-specific parameters $\bs\beta_1,\ldots, \bs\beta_m$  
together
through the parameters $(\bs\beta_0,\Psi)$. 
For given values of
$(\bs\beta_0, \Psi,\sigma_1^2,\ldots, \sigma^2_m)$,
the mean squared estimation 
error for each ${\bs\beta}_j$, on average with respect to  (\ref{eqn:agm}), 
is minimized by the conditional expectation 
$\tilde{\bs\beta}_j =  E( \bs \beta_j | \blt y_j) =  (\bs \Psi^{-1} + \blt X_j^\top \blt X_j /\sigma^2_j   )^{-1} 
   ( \blt X_j^\top \blt y_j/\sigma^2_j + \Psi^{-1} \bs\beta_0 )$.
This estimator can be interpreted as a Bayes estimator when 
(\ref{eqn:agm}) is thought of as a prior distribution
\cite{Lindley1972} and is sometimes 
referred to as the ``best linear unbiased predictor'' (BLUP) 
when (\ref{eqn:agm}) is thought of as a sampling model for the groups \cite{Henderson1975},
despite the fact that $\tilde{\bs\beta}_j$ is biased as an estimator 
of ${\bs\beta}_j$ (the ``U'' in ``BLUP'' refers to 
average bias with respect to (\ref{eqn:agm}) - essentially the across-group average 
bias of $\tilde{\bs\beta}_1,\ldots, \tilde{\bs\beta}_m$). 
In practice, values of 
$(\bs\beta_0,\Psi,\sigma^2_1,\ldots, \sigma^2_m)$ are estimated 
from $\blt y_1,\ldots, \blt y_m$ and then 
plugged into the 
equation for each $\tilde{ \bs\beta}_j$, yielding a so-called 
empirical Bayes estimator that is an ``indirect estimator'' in the 
sense that it
combines direct information from group $j$ 
with indirect information from other groups, via the estimates 
of $(\bs\beta_0,\Psi,\sigma^2_1,\ldots,\sigma^2_m)$.
While 
the empirical Bayes estimator for a given group $j$ has 
potentially higher mean squared error than the corresponding OLS estimator, 
the across-group average mean squared error of the empirical 
Bayes estimators is 
typically lower than that of the 
OLS estimators, 
even if the linking model (\ref{eqn:agm}) is incorrect
or only a conceptual device - for example if the groups are not 
randomly selected. 

Analogously, we may consider direct and indirect methods for 
hypothesis testing. The most widely used frequentist level-$\alpha$ 
direct test of a 
linear hypothesis about  $\bs\beta_j$ is the standard $F$-test, 
whose test statistic is a function of data only from group $j$. 
In contrast, a frequentist level-$\alpha$ indirect test is one where the 
test statistic for group $j$ 
is allowed to depend on the data from the other groups. 
In particular, data from groups other than $j$ might suggest that 
the vector ${\bs\beta}_j$ lies in a particular direction.
We can use this indirect information to select a level-$\alpha$ test of 
$\bs\beta_j$ that has more power than the $F$-test in this 
particular direction, at a cost of having less power in other directions. 
As long as the 
data are independent across-groups, such a procedure will 
maintain a type I error rate of $\alpha$ if the 
hypothesis is true, while having having increased power 
as compared to the $F$-test if the hypothesis is false 
and the indirect information from the other groups is reasonably accurate. 

In comparison to methods for 
indirect estimation, methods for indirect hypothesis 
testing are relatively undeveloped. Notable work in this area by O'Gorman \cite{OGorman2002,OGorman2006} examines adaptive procedures for testing subsets of regression coefficients in a linear model. When the errors are non-normal, such procedures have higher power than the $F$-test for a variety of error distributions. Recently, Hoff \cite{Hoff2021}
proposed an indirect analogue to the standard $t$-test and 
corresponding $p$-value for a univariate
parameter $\theta_j$ based on a normally distributed estimator 
$\hat\theta_j$ and indirect data from groups other than group $j$. 
The indirect test of $H_j:\theta_j=0$ 
has a rejection region that is asymmetric around zero and is chosen 
to maximize expected power with respect to a ``prior distribution'' 
that is derived from indirect data that is independent of $\hat\theta_j$. 
Such a test is ``frequentist'', as it maintains an exact level-$\alpha$ 
type I error rate, but is also Bayesian in that it minimizes a 
Bayes risk (one minus the prior expected power) and so it 
is referred to as being ``frequentist and Bayesian'', or FAB. 
Inversions of such tests were used by Yu and Hoff \cite{Yu2018}
to construct indirect confidence intervals 
for means in multiple normal populations. Their confidence intervals 
are essentially a multigroup extension of the interval proposed by  Pratt \cite{Pratt1963}
who obtained 
the confidence interval for the mean of a normal population
that has minimum prior expected width among those having $1-\alpha$ 
frequentist coverage. These Bayes-optimal frequentist 
procedures can be derived 
from various likelihood ratios, and so 
in this sense, they are related to methods that use 
Bayes factors as statistics in frequentist hypothesis 
tests \cite{Chacon2007,Good1992}. Such an approach has been applied to 
testing hypotheses about multinomial probabilities
\cite{Good1974}
and for evaluating nonparametric goodness of fit
\cite{Aerts2004}.

In this article we develop FAB alternatives to the 
$F$-test for evaluating group-specific linear hypotheses in multigroup 
regression settings. Essentially, the proposed FAB test 
for $\bs\beta_j$ is a level-$\alpha$ test that has maximum 
expected power with respect to the ``prior'' distribution 
$\bs\beta_j \sim N_p( \hat{\bs\beta}_0 ,\hat{\Psi})$, where 
$(\hat{\bs\beta}_0 ,\hat\Psi)$ are estimated from models 
(\ref{eqn:wgm}), (\ref{eqn:agm}) and data from groups other than $j$. 
In the next section, we derive the form of the optimal FAB statistic and obtain a numerical 
approximation to facilitate its calculation. 
A theoretical power comparison of a simplified version of the FAB test, which we call the cone test, and the $F$-test appears in Section 3. It is shown 
that the $F$-statistic is a special case of the FAB statistic. It is also shown that the ratio of the $F$-test $p$-value 
to the cone test $p$-value may range from zero to infinity. Asymptotic power comparisons between the $F$-test and the cone test are also provided, where it is shown that the cone test can have higher power than the $F$-test in certain high-dimensional scenarios. In Section 4 
we describe 
in greater detail how the FAB statistic can be used in multigroup settings, and review some methods for obtaining estimates of the linking model parameters. A data analysis example considering educational test scores from multiple schools appears in Section 5. The FAB tests that share information across schools 
lead to a substantially greater number of null hypotheses being rejected. 
A discussion follows in Section 6. All of the proofs of theoretical results presented in this article can be found in the Appendix.

\section{A FAB test for linear hypotheses} 
\label{Sec:BasicFABTest}

Consider testing a linear hypothesis 
for the parameter $\bs\beta\in \mathbb R^p$ 
based on an observation of $\blt y$ from the 
linear model
$\blt y\sim N_n(\blt X\bs\beta,\sigma^2\blt I)$, where $\blt X\in\mathbb R^{n\times p}$ is a known 
matrix of predictors and $\bs\beta$ and $\sigma^2$ are unknown. For the moment it is assumed that $\blt X$ has full-rank. 
Recall that 
linear hypotheses of the form $H: \blt A \bs\beta = \blt c$ 
may be 
expressed as $H: \bs\theta = \blt 0$ for a transformed linear model with mean 
$ \tilde{\blt X} \bs\theta$ 
(Seber \& Lee, 2003)
so without loss of generality 
we consider the null hypothesis $H: \bs\beta=\blt 0$. 
Note that the null  and alternative models are 
invariant under data rescalings of the form $\blt y \rightarrow c \blt y$ for
scalars $c\neq 0$.  
For this reason, 
we restrict attention to tests based on statistics
that are invariant with respect to the group of rescalings of $\blt y$. 
Any such statistic must be a function of a maximal invariant statistic such as the unit vector $\blt u=\blt y/||\blt y||$. 
A test statistic based on $\blt u$ has the advantage that its
null distribution does not depend on any unknown parameters, 
as $\blt u$ is uniformly distributed on the sphere $\mathbb S^{n-1}$ 
if $\bs\beta =  \blt 0$.

One scale invariant statistic is the 
usual $F$-statistic, 
$F(\blt y) = \{  (SST - SSR)/p \}/$ $\{  SSR/(n-p) \} $
where $SST = \blt y^\top \blt y$ is the total sum of squares
and $SSR = ||\blt y-\blt X\hat{\bs\beta}||^2$ is 
the residual sum of squares, with $\hat{\bs\beta}$ being the
OLS estimator. To see that this depends on $\blt y$ only through $\blt u$, 
let $\blt P $
be the projection matrix onto the space spanned by the
columns of $\blt X$. Then we can write
\begin{align*}
F(\blt y) & =  \frac{n-p}{p} \frac{ \blt y^\top \blt y - \blt y^\top (\blt I- \blt P) \blt y }{\blt y^\top 
           (\blt I-\blt P)\blt y } \label{eqn:fstat}  \\
 &= \frac{n-p}{p} \frac{ \blt y^\top\blt  P\blt  y/\blt  y^\top \blt y }{ 1-\blt  y^\top \blt P\blt  y/\blt y^\top\blt  y } =
  \frac{n-p}{p} \frac{ \blt u^\top \blt P \blt u}{ 1- \blt u^\top \blt P \blt u}. 
 \nonumber
\end{align*} 
Note that 
If $n\leq p$ then  $\blt P=\blt I$ and $\blt u^\top \blt P \blt u =1$ is constant in $\blt y$, and so in this case
any test based on $\blt u^\top\blt  P\blt  u$ (such as the $F$-test) has power
equal to its level. Even if $n>p$, since the 
distribution of 
the $F$-statistic depends on 
$\bs\beta$ only through $||\blt X \bs\beta||^2$, 
its power is constant on level-sets of $||\blt X\bs\beta||^2$.
In this sense the $F$-test is ``looking'' in all directions  
equally for evidence against the null hypothesis. 

If prior information about the direction of $\bs\beta$ is available, 
it may be preferable to use a test that has more power in this direction,
at the cost of
having lower power in the opposite direction. 
Specifically, suppose prior information about $(\bs\beta,\sigma^2)$
is available
in the form of a prior density $\pi(\bs\beta,\sigma^2)$. 
Letting $p(\blt u | \bs\beta,\sigma^2)$ be the density 
of $\blt u$ implied by the normal model
(\ref{eqn:wgm}). The
prior expected power of a test function $\phi: \mathbb S^{n-1} \rightarrow [0,1]$ 
is given by
\begin{align*}
E( \phi )  & =  
\int \int \int \phi(\blt u) \, p(\blt u|\bs\beta,\sigma^2) \pi(\bs\beta,\sigma^2)    \, d\blt u \, d\bs\beta \, d\sigma^2   \\
  & = \int \phi(\blt u) \, p_\pi(\blt u) \, d\blt u,
\end{align*} 
where $p_\pi(\blt u)= \int p(\blt u|\bs \beta,\sigma^2) \,  \pi(\bs\beta,\sigma^2) \, d\bs \beta \, d\sigma^2$ is the marginal density density of $\blt u$ induced by $\pi$, with respect to the uniform measure on $\mathbb S^{n-1}$. 
The Bayes-optimal
level-$\alpha$ test $\phi_{\pi}$ is the test that maximizes $E(\phi)$
among all level-$\alpha$ tests. 
Since the null distribution of $\blt u$ is uniform on $\mathbb S^{n-1}$, 
by the Neyman-Pearson lemma the optimal test is given by
\begin{align*}
  \phi_\pi(u)  = \left \{  \begin{array}{r} 
     1 \ \text{if $p_\pi(\blt u)>c_\alpha$ } \\
     0 \ \text{if $p_\pi(\blt u)\leq c_\alpha$ }  
       \end{array} \right  .
\end{align*}
where $c_\alpha $ is chosen so that  $\phi_\pi$ 
has a type I error rate equal to $\alpha$. 

Primarily for computational
reasons we consider the case that 
 $\pi$  corresponds to a 
normal distribution 
$\bs\beta\sim N_p(\bs\beta_0,\Psi)$ for $\bs\beta$
and a point-mass distribution on $\sigma^2_0$ for  $\sigma^2$. 
The marginal distribution of $\blt y$ under this prior 
is $\blt y \sim N_n( \blt X \bs\beta_0, \blt X \Psi \blt X^\top + \sigma^2_0 \blt I)$, 
and the corresponding distribution for $\blt u$ is 
the angular Gaussian distribution 
$\blt u \sim AG(\bs \mu ,\Sigma)$ with 
$\bs\mu = \blt X \bs \beta_0 $ 
and $\Sigma= \blt X \Psi \blt X^\top +  \sigma^2_0 \blt I$. 
The density of $\blt u\sim AG(\bs\mu,\Sigma)$ 
with respect to the uniform
probability distribution on the sphere
is derived in Pukkila \& Rao (1988)
and is 
 given by
 $p(\blt u|\bs\mu,\Sigma) \propto x^{-n}  e^{r^2/2} I_n(r) $
where $x= \sqrt{ \blt u^\top\Sigma^{-1}\blt  u }$, 
$r=\blt  u^\top\Sigma^{-1}\bs \mu /x$, and $I_n(r)=
\int_0^\infty z^{n-1} e^{-(z-r)^2/2} \, dz$, approximation and 
computation of which is 
described below. 
 Therefore, the Bayes-optimal test of $H:\bs\beta=\blt 0$ based on 
$\blt u$
rejects the hypothesis when $x^{-n}  e^{r^2/2} I_n(r) $ is large, 
or equivalently, for large values of 
the FAB test statistic
\begin{equation}
 T_{FAB}(\blt u) = r^2/2 + \log I_n(r) - n \log x.  
\label{eqn:tfab}
\end{equation}
The function 
$I_n(r)$
 can be computed recursively as 
\begin{align*} 
I_1(r) &= 2\pi \Phi(r) \\
I_2(r) &= e^{-r^2/2} + r I_1(r) \\
I_n(r) &= (n-2) I_{n-2}(r) + r I_{n-1}(r), 
\end{align*}
where $\Phi$ is the standard normal CDF \cite{Pukkila1988}.
While the recursion can be performed quite quickly, 
it can be numerically unstable if 
$r$ is negative and $n$ is large. 
Alternatively, for large $n$ we have the following approximation:
\begin{align*}
    \log   I_n(r) \approx (n/2-1)\log 2 + \log \Gamma( n/2) + 
    \sqrt{n} r - r^2/4. 
\end{align*}
This is based on a Taylor series expansion and then applying
Rocktaeschel's \cite{RockApprox} approximation to the gamma function.
Based on this approximation, an approximately optimal test 
statistic is 
\begin{equation}
T_{AFAB}(\blt u)  =  r^2/4+ \sqrt{n} r - n \log x. 
\label{eqn:tafab} 
\end{equation}
The null distributions of $T_{FAB}$ and $T_{AFAB}$ may be 
easily obtained via Monte Carlo simulation, because 
under the null hypothesis $H:\bs\beta=\blt 0$, the distribution of
$\blt u$ is uniform on $\mathbb S^{n-1}$ and so it is
free of any unknown 
parameters.
Furthermore, an approximation to the 
 $p$-value corresponding to any test statistic $T$ based on 
$\blt u = \blt y/\Vert \blt y\Vert$ 
may be obtained as follows:

\begin{enumerate}
\item  Simulate  $\blt y^{(1)} ,\ldots, \blt y^{(S)} \sim \text{i.i.d. }  N_n(\blt 0,\blt I)$;
\item Compute $\blt u^{(s)} = \blt y^{(s)} /\Vert \blt y^{(s)} \Vert$ for 
 $s=1,\ldots, S$;
\item  Compute $\hat p = \tfrac{1}{S} \sum_{s=1}^S I\big(  T(\blt u^{(s)}) \geq T(\blt u) \big).$ 
\end{enumerate}

This Monte-Carlo approximation $\hat p$ to the
 actual $p$-value
can be made arbitrarily accurate by increasing the Monte Carlo 
sample size $S$. 

We remind the reader that 
the power of the test statistics $T_{FAB}$ and $T_{AFAB}$ depend on 
$\bs\beta$, the normality of $\blt y$ and the 
prior distribution $\pi$, 
but the frequentist validity of the tests depend only 
on the distribution of $\blt u$ being uniform under the null 
hypothesis. This means that the tests described above 
are also valid for testing $H:\bs\beta=\blt 0$ in any linear regression 
model $\blt y = \blt X\bs \beta + \sigma \bs\epsilon$ where the 
distribution of $\bs\epsilon$ is spherically symmetric. 
This includes non-Gaussian heavy-tailed error distributions, 
such as the $t$ and Cauchy distributions. 
If spherical symmetry under the null distribution is 
suspect, an alternative technique would be to
 obtain 
the null distribution of $\blt u$ via randomization. Such a randomization 
test may be obtained by replacing 
$\blt y^{(1)} ,\ldots, \blt y^{(S)}$ in 
 step 1  above 
with $S$ independent permutations of the elements of the observed data 
vector $\blt y$.

Another desirable feature of the FAB test is that, unlike the $F$-test, when $n \leq p$ the FAB test has non-trivial power against certain alternative hypotheses, meaning that the power of the test is strictly greater than the level of the test. In contrast, the $F$-test has power equal to the level of the test when $n \leq p$. If the null hypothesis is false and the FAB prior distribution is concentrated around the true parameter values, the FAB test will have non-trivial power. This is especially useful in settings where the sample size is small or when there are a large number of predictor variables.  Conceptually, when $n  \leq p$ the FAB test is comparable to ridge regression, where a prior distribution is used to add additional structure to an otherwise degenerate inference problem. Regardless of the rank of $\blt X$, both the $F$ and FAB tests maintain the correct level.  It should be noted that if $\blt X$ is not full-rank, there does not exist an unbiased test of the hypothesis $H:\bs \beta = \blt 0$ due to the lack of identifiability of the regression model. In the next section we provide some additional insight on the performance of the FAB test relative to the $F$-test as a function of the values of $n$ and $p$.

\section{Theoretical power comparisons}
\label{Sec:Theory}
 In this section we first examine the FAB test statistic under the prior distribution $\bs \beta \sim N_p(\bs \beta_0,\gamma (\blt X^\top \blt X)^{-1})$. In the extreme cases where either $\bs \beta_0 = \blt 0$ or $\gamma \rightarrow \infty$ the FAB test is equivalent to the $F$-test. In another extreme case where $\gamma = 0$, the FAB test is equivalent to a test which we call the cone test. When both $\bs \beta_0$ and $\gamma$ are non-zero the FAB test can be seen as an interpolation between the simpler $F$ and cone tests. As such, in the remainder of this section we compare the asymptotic performance of the $F$ and cone tests. In particular, it will be shown that the cone test, if correctly specified, can significantly outperform the $F$-test when the dimension $p$ of the regression subspace is large. This suggests that the FAB test is especially useful in settings with a large number of predictor variables, in which the $F$-test has low power.

\subsection{The FAB, F and cone tests}
Suppose that the prior distribution $\bs \beta \sim N_p(\bs \beta_0,\gamma (\blt X^\top \blt X)^{-1})$ is used in the FAB test statistic \eqref{eqn:tfab}. Under this prior distribution, the marginal distribution of $\blt X \bs \beta$ is $N(\blt X \bs \beta_0,\gamma \blt P)$, where $\blt P$ is the orthogonal projection matrix onto $\text{col}(\blt X)$. This marginal distribution has the property that $E(\blt X \bs \beta) = \blt X \bs \beta_0$ and the distribution of $\blt X \bs \beta - \blt X \bs \beta_0$ is rotationally invariant for all orthogonal transformations that fix the subspace $\text{col}(\blt X)$. This class of prior distributions with covariance matrices of the form $\gamma (\blt X^\top \blt X)^{-1}$ is useful in situations where it is not feasible to model the entire prior covariance matrix of $\bs \beta_0$. Moreover, intuition for the behaviour of the FAB test statistic can be obtained by using this class of prior distributions. We have that $\bs \Sigma^{-1} = (\gamma \blt P + \sigma_0^2\blt I)^{-1} = \sigma_0^{-2}(\blt I - \frac{\gamma}{\gamma + \sigma_0^2}\blt P) = \sigma_0^{-2}\big((1-w)\blt I + w(\blt I  - \blt P)\big)$ where $w = \gamma/(\gamma + \sigma_0^2)$. Expanding the expression \eqref{eqn:tfab}, the FAB test statistic in this case takes the form
\begin{eqnarray}
    T_{FAB}(\blt u) =  n\log \big( \sigma_0\big) - \frac{n}{2} \log\big(1 - w + w\Vert (\blt I - \blt P)\blt u\Vert^2\big) + \nonumber
    \\
     \log\bigg(\int_0^\infty z^{n-1}\exp\big(-z^2/2 + z\frac{ (1-w) \blt u^\top \blt X \bs\beta_0}{(1-w + w \Vert (\blt I - \blt P)\blt u\Vert^2)^{1/2}} \big)dz \bigg). \label{eqn:FABStatSimplified}
\end{eqnarray}
The last term in \eqref{eqn:FABStatSimplified} is simply an expansion of the term $\log I_n(r)$ in \eqref{eqn:tfab}. This expression for the FAB test statistic shows that $T_{FAB}$ is a function of $\Vert (\blt I - \blt P)\blt u\Vert^2$ and $\blt u^\top \blt X \bs \beta_0$. For a fixed value of $\Vert (\blt I - \blt P)\blt u\Vert^2$, $T_{FAB}$ is a strictly increasing function of $\blt u^\top \blt X \bs \beta_0$, while for a fixed, positive value of $\blt  u^\top \blt X \bs \beta_0$, $T_{FAB}$ is a strictly decreasing function of $\Vert (\blt I - \blt P)\blt u \Vert^2$. Consequently, the FAB test statistic is large when $\blt u$ is simultaneously close to the vector $\blt X \bs \beta_0/\Vert \blt X \bs \beta_0 \Vert$  and close to the subspace $\text{col}(\blt X)$. From \eqref{eqn:FABStatSimplified} it is immediate that when $\bs \beta_0 = \blt 0$, $T_{FAB}$ is a strictly decreasing function of $\Vert (\blt I - \blt P)\blt u \Vert^2$ and so it results in a test that is equivalent to the $F$-test. Similarly, if $\gamma \rightarrow \infty$ the prior distribution becomes diffuse and the FAB test statistic converges pointwise to a limiting test statistic that is a strictly decreasing function of $\Vert (\blt I - \blt P) \blt u\Vert^2$. It is shown in the Appendix that the FAB test is asymptotically equivalent to the $F$-test as $\gamma \rightarrow \infty$.

Another extreme case occurs when $\gamma = 0$ so that $w = 0$ and $T_{FAB}$ is a strictly increasing function of $\blt u^\top \blt X  {\bs \beta}_0$. We call the resulting test with $\gamma = 0$ the cone test with test direction $\bs \mu = \blt X \bs \beta_0/\Vert \blt X \bs \beta_0 \Vert \in \mathbb S^{n-1}$.  When the level of the cone test is less than $1/2$, the test has the rejection region 
\begin{equation}
\label{eqn:coneRejR}
    R_{\alpha} = \bigg\{\blt y: \frac{\blt y^\top \bs \mu}{\Vert \blt y \Vert} >  c_{1 - \alpha}\bigg\}, \;\; \bs \mu \in \mathbb S^{n-1},
\end{equation}
where the number $c_{1-\alpha} > 0$ is chosen to make this a level-$\alpha$ test. We call the test with rejection region $R_\alpha$ a cone test because the set $R_\alpha$ forms a cone in $\mathbb R^n$ that is rotationally symmetric about the the ray extending in the direction $\bs \mu$ from the origin. By construction, the cone test is identical to the likelihood ratio test of $H:\bs \beta = \blt 0$ against the simple alternative hypothesis $H_a: \bs \beta = \bs \beta_0$.

\subsection{Power and $p$-value comparisons of the $F$ and cone tests}
The feature of the test statistic \eqref{eqn:FABStatSimplified} that differentiates the FAB test from the $F$-test is the dependence of the FAB test statistic on $\blt u^\top \blt X \bs \beta_0$. The quantity $\blt u^\top \blt X \bs \beta_0$ is the cosine of the angle between the scaled data $\blt y/\Vert \blt y \Vert$ and the direction of the cone test. As the prior distribution becomes more concentrated about $ \bs \beta_0$ the FAB test can be approximated by the cone test. Below we compare the asymptotic properties of the $F$-test and the cone test with rejection region given by \eqref{eqn:coneRejR}.

Our first lemma compares the $p$-value functions of the $F$-test and cone test. In the univariate setting, it is shown in 
Hoff \cite{Hoff2021} that for a given observation $y$, the $p$-value function of the $F$-test $p_F(y)$ (or equivalently the $p$-value function of the two-sided $t$-test) at $y$ can be at most twice as large as the $p$-value function of the FAB test $p_{FAB}(y)$. However, the ratio $p_{F}(y)/p_{FAB}(y)$ can be arbitrarily close to $0$, meaning that it is possible for the FAB $p$-value to be significantly larger than the $F$-test $p$-value if the true $\bs \beta$ points in a direction opposite to that of $\bs \beta_0$. The following lemma shows that in a multivariate setting the $p$-value ratio $p_F(\blt y)/p_{FAB}(\blt y)$ instead of being bounded above by two, is also unbounded. In particular, for observations $\blt y$ where $\blt P \blt y  = \lambda \bs \mu ,\; \lambda > 0$, the $p$-value ratio of such an observation can be expressed as a ratio of probabilities of Dirichlet random variables. 

\begin{lemma}{Lemma 1}{}
\label{Lemma:PvalueLemma}
Let $p_C(\blt y)$ be the $p$-value function for the cone test with rejection region \eqref{eqn:coneRejR} for testing the null hypothesis $H_0:\bs \beta =\blt 0$ where $\bs \mu \in \text{col}(\blt X)$ with $\Vert \bs \mu \Vert = 1$. If the observation $\blt y$ is of the form $\blt y = a\bs \mu + b \blt v$ with $\blt v \in \text{col}(\blt X)^\perp$, $a,b > 0$ then
\begin{align*}
    \frac{p_F(\blt y)}{p_C(\blt y)} = \frac{P\big( \sum_{i = 1}^p s_i^2 > c(\blt y) \big)}{P\big(s_1^2 > c(\blt y)\big)},
\end{align*}
where $(s_1^2,\ldots,s_n^2) \sim \text{Dirichlet}_n(\frac{1}{2},\ldots,\frac{1}{2})$ and $c(\blt y) = a^2/(a^2 + b^2)$.
In particular, for such a $\blt y$ the $p$-value ratio can be bounded below by
\begin{align*}
\frac{p_F(\blt y)}{p_C(\blt y)} \geq \frac{4}{(n-p)} \big(\frac{c(\blt y)}{1-c(\blt y)}\big)^{\frac{p-1}{2}} ,
\end{align*}
which tends to $\infty$ as $c(\blt y) \rightarrow 1$ if $1 < p < n$. 
\end{lemma}

From this lemma it is seen that for any observation $\blt y$ with $\blt P \blt y  = \lambda \bs \mu ,\; \lambda > 0$, the $p$-value of the $F$-test is larger than the $p$-value of the cone test.  Moreover, for such an observation, the $p$-value ratio converges to infinity as $\Vert (\blt I - \blt P) \blt y \Vert \rightarrow 0$ when $p > 1$. Therefore, unlike in the univariate setting, the $p$-value of the cone test can be orders of magnitude smaller than the corresponding $F$-test $p$-value when $\blt y$ is close to $\bs \mu$. In practice, it is not realistic to observe a $\blt y$ with $\blt P \blt y  = \lambda \bs \mu ,\; \lambda > 0$. However, since the $p$-value ratio $p_F(\blt y)/p_C(\blt y)$ is continuous at $\blt y$ as long as $p_C(\blt y) \neq 0$, the conclusion of Lemma 1 can be extended to observations $\blt y$ with  $\blt P \blt y  \approx \lambda \bs \mu ,\; \lambda > 0$. For instance, if $p_F(\blt y)/p_C(\blt y) > M$ then there is a neighbourhood of $\blt y$ where this inequality holds for $\blt y$ within this neighbourhood.

As the cone test only depends on the regression subspace $\text{col}(\blt X)$ through $\bs \mu \in \text{col}(\blt X)$, the performance of the cone test is independent of the dimension $p$ of the regression subspace. This contrasts with the power of the $F$-test which deteriorates as $p$ grows. Formalizing this, consider the sequence of models
\begin{equation}
    \mathcal{M}_n:\blt y_n \sim N_n(\blt X_n \bs \beta_n,\sigma^2\blt I), \;\; \bs \beta_n \in \mathbb{R}^{p_n}, n \in \mathbb{N}.
    \label{eqn:ModelSequence}
\end{equation}
 Define $\rho_n(c_n,p_n,\sigma^2)$ to be the power of a level-$\alpha$ test ($\rho_n$ will be the power of either the $F$-test or the cone test) of the null hypothesis $H_{0,n}:\bs \beta_n = \blt 0$ under the alternative hypothesis that has $\Vert \blt X_n \bs \beta_n\Vert = c_n$. It is of interest to assess the impact of $p_n$ and $n$ on the power functions of the $F$ and cone tests. Table 1 summarizes the limiting power $ \liminf_{n \rightarrow \infty}\rho_n(c_n,p_n,\sigma^2)$ of both the $F$-test and the cone test under various asymptotic regimes, when the test direction of cone test is correctly specified. The cone test direction is correctly specified when the null hypothesis does not hold and $\blt X \bs \beta/ \Vert \blt X \bs \beta \Vert = \bs \mu$. These asymptotic regimes differ based on whether $p_n$ is taken to be fixed as $n$ increases or $p_n/n \rightarrow \gamma \in (0,1)$ as $n$ increases. Summarizing Table 1, the correctly specified cone test will have higher limiting power than the $F$-test in settings with a large number of predictor variables. 
\begin{table}[ht]
\centering
\begin{tabular}{|r|rr|}
  \hline
 & $F$-test & Cone test \\ 
  \hline
  \hline
 $p_n = p_0$, $c_n \rightarrow \infty$ & $\rho^* = 1$ &  $\rho^* = 1$ \\
$p_n = p_0$, $c_n = c_0$ & $\alpha < \rho^* < 1$ & $\alpha < \rho^* < 1$ \\ 
 $p_n = \lfloor \gamma n \rfloor$, $c_n = c_0$ & $\rho^* = \alpha$  & $ \alpha < \rho^* < 1$  \\ 
 $p_n = \lfloor \gamma n \rfloor$, $c_n = n^{1/4}$ & $\alpha < \rho^* < 1$ & $\rho^* = 1$ \\
   \hline
\end{tabular}
\label{Tab:AsymptoticPvalues}
\caption{Limiting power $\rho^* = \liminf_{n \rightarrow \infty}\rho_n(c_n,p_n,\sigma^2)$ of the $F$-test and the cone test with test direction $\bs \mu = \blt X_n \bs \beta_n/\Vert \blt X_n \bs \beta_n \Vert$, where $p_0$ and $c_0$ are constants.}
\end{table}
A more complete description of the asymptotics of the $F$-test is provided in the following lemma. This lemma shows that under the regime where $p_n/n \rightarrow \gamma$ as $n \rightarrow \infty$, the value of $\Vert \blt X_n \bs \beta_n \Vert$ needs to diverge from $0$ at a rate of $\Vert \blt X_n \bs \beta_n \Vert = n^{1/4}$ if the power of the $F$-test is to be greater than its level. 

\begin{lemma}{Lemma 2}{}
\label{Lemma:FTestAsympt}
Let $\rho_n(c_n,p_n, \sigma^2)$ denote the power of the level-$\alpha$ $F$-test in the sequence of models \eqref{eqn:ModelSequence}. If $p_n = p_0$ and $c_n = c_0$ are constants then $\alpha < \liminf_{n \rightarrow \infty} \rho_n(c,p,\sigma^2) < 1$. If $\gamma \in (0,1)$ then  $\liminf_{n \rightarrow \infty} \rho_n(c_0,\lfloor \gamma n\rfloor,\sigma^2)  = \alpha$ and if $c_n = n^{1/4}$ then the $F$-test has limiting power $\liminf_{n \rightarrow \infty} \rho_n(n^{1/4},\lfloor \gamma n \rfloor,\sigma^2) \in (\alpha,1)$.    
\end{lemma}

Similarly, the results in Table 1 regarding the cone test are described in the following lemma, where these results are extended to cone tests that are not correctly specified.    

\begin{lemma}{Lemma 3}{}
\label{Lemma:coneAsymp}
Let $\rho_n(\bs \mu_n,\blt v_n, p_n, \sigma^2)$ denote the power of the level-$\alpha$ cone test with rejection region $\{\blt y:\langle \blt y/\Vert \blt y\Vert,\bs \mu_n\rangle > q_{n,1-\alpha}\}$ in the sequence of models \eqref{eqn:ModelSequence}, where $q_{n,1-\alpha}$ is an appropriate level-$\alpha$ quantile, $\bs \mu_n \in \mathbb{S}^{n-1}$ and $\blt v_n = \blt X_n \bs \beta_n$. If $\Vert \blt v_n\Vert^2  = c_0$ is constant and the mean direction of the cone test is nearly correctly specified so that $\Vert \bs \mu_n - \blt v_n/c_0 \Vert = o(n^{-1/2})$ then $\liminf_{n \rightarrow \infty} \rho_n(\blt v_n, \bs \mu_n, p_n, \sigma^2) \in (\alpha,1)$ where the power function does not depend on $p_n$. If $\Vert \blt v_n\Vert = n^{1/4}$ and if $\Vert \bs \mu_n - (\blt v_n/\Vert \blt v_n \Vert) \Vert = (n^{-1/4} - an^{-1/2})$ for some $a > 0$ then  $\liminf_{n \rightarrow \infty} \rho_n(\bs \mu_n, \blt v_n, p_n, \sigma^2) \in (\alpha,1)$ and if $\Vert \bs \mu_n - (\blt v_n/\Vert \blt v_n \Vert) \Vert = o(n^{-1/4})$ then  $\liminf_{n \rightarrow \infty} \rho_n(\bs \mu_n, \blt v_n, p_n, \sigma^2) = 1$.
\end{lemma}

This lemma implies that the performance of a mispecified cone test depends on the dimension $n$, where the test direction $\bs \mu_n$ must converge to the direction of the mean vector under the alternative $\blt X_n \bs \beta_n/\Vert \blt X_n \bs \beta_n \Vert$ at a rate of $o(n^{-1/2})$ if the test is to have limiting power that is greater than $\alpha$. If $\theta_n = \angle (\blt 0,\blt v_n,\blt X_n \bs \beta_n)$ is the angle between $\blt v_n$ and $\blt X_n \bs \beta_n$ then this condition is equivalent to $\theta_n = o(n^{-1/2})$. Essentially, if $\theta_n = \theta_0 > 0$ is constant, the power of the cone test diminishes as $n$ increases but does not depend on $p_n$, while the power of the $F$-test diminishes as $p_n$ increases but it is not significantly altered by $n$.

In both lemmas 2 and 3, it is not unrealistic to assume that $\Vert \blt X_n \bs \beta_n \Vert \rightarrow \infty$ when the null hypotheses $H_{0,n}$ do not hold. One common such scenario is where $m$ independent replications are observed, each with design matrix $\tilde{\blt X}, \in \mathbb{R}^{k \times p_0}$ so that, $p_n = p_0$ and $\bs \beta_n = \bs \beta_0$ are constant while  $\blt X_{n} = (\blt 1_{m} \otimes  \tilde{\blt X}),\; n = km$. In this case $c_n = \Vert \blt X_n \bs \beta_n \Vert = \sqrt{n/k} \Vert \tilde{\blt X} \bs \beta_0 \Vert$ and the $F$ and correctly specified cone tests will have limiting power $1$ by the above lemmas. A related case is when $p_n = p_0$, $\bs \beta_n = \bs \beta_0$ for all $n$ and the entries of $\blt X_n$ are independent standard normal random variables. Then $E(\Vert \blt X_n \bs \beta_n \Vert^2) = n \Vert \bs \beta_0 \Vert^2$ and $E(\Vert \blt X_n \bs \beta_n \Vert) \asymp \sqrt{n}$. Therefore, the rate $c_n = n^{1/4}$ appearing in lemmas 2 and 3 has the alternative hypothesis diverging away from the null hypothesis at a rate that is slower than the $\sqrt{n}$-rate that occurs in the above  replication scenarios.

It is not realistic to assume that the cone test direction is correctly specified in practice. In the multigroup setting to be considered in the following section, even as the amount of auxiliary information from other groups increases, it will generally not be true that the test direction of the cone test can be estimated consistently from the auxiliary information. Therefore we do not recommend using the cone test on its own. Rather, we recommend using the FAB test statistic \eqref{eqn:tfab} or \eqref{eqn:tafab}, which provides a principled compromise between the $F$ and cone tests. If the prior information about $\bs \beta$ is precise and accurate, the resulting FAB test will look similar to the cone test. If instead only weak prior information is available, the FAB test will behave similarly to the $F$-test.   

\section{FAB testing in multigroup settings}
\subsection{The multigroup FAB Test}
Thus far, the prior distribution $\bs \beta \sim N_p(\bs \beta_0,\bs \Psi)$ used in the FAB test was assumed to known at the outset. In this section we demonstrate how to choose the prior distribution in a data dependent manner in a multigroup setting.  Specifically, we consider the model \eqref{eqn:wgm} where $\blt y_j \in \mathbb R^{n_j}$, and $\bs \beta_j \in \mathbb R^p$.  We construct $m$ different FAB tests for the each of the separate hypotheses $H_j:\bs \beta_j = \blt 0, j = 1,\ldots,m$ in this model. To ease notation, we focus on testing the hypothesis $H:\bs \beta_1 = \blt 0$, where any of the other hypotheses can be tested in a similar manner by relabelling the groups.

Given the multigroup regression model \eqref{eqn:wgm}, the prior distribution in \eqref{eqn:agm}, which assumes that the regression coefficients are drawn from a common multivariate normal distribution, is used as a device to share information across the different groups in our multigroup FAB test. Such a prior distribution will also be referred to as a linking model. If the parameters $\bs \beta_0$ and $\bs \Psi$ in the linking model \eqref{eqn:agm} were known, the FAB test introduced in Section \ref{Sec:BasicFABTest}
could be directly applied. As this is generally not the case, the data from groups $2$ through $m$ will be used to obtain estimates of $\bs \beta_0$ and $\bs \Psi$. A level-$\alpha$ multigroup FAB procedure can be constructed as follows:
\begin{enumerate}
    \item Obtain estimates $\hat{\bs \beta}_0 = \hat{\bs \beta}_0 (\blt y_2,\ldots,\blt y_m)$ and $\hat{\bs \Psi} = \hat{\bs \Psi}(\blt y_2,\ldots,\blt y_m)$ of the linking model parameters using observations from every group but the first group.
    \item Plug-in the values of $ \hat{\bs \beta}_0$ and $ \hat{\bs \Psi}$ into the FAB test statistic in \eqref{eqn:tfab}, where $\bs \mu = \blt X_1  \hat{\bs \beta}_0$ and $\bs \Sigma = \blt X_1  \hat{\bs \Psi} \blt X^\top + \sigma_0^2 \blt I$. Denote the observed FAB test statistic by $ T_{FAB}\big( \frac{\blt y_1}{ \Vert \blt y_1 \Vert}, \hat{\bs \beta}_0,\hat{\bs \Psi}\big)$.
    \item Reject the null hypothesis if $ T_{FAB}\big( \frac{\blt y_1}{ \Vert \blt y_1 \Vert}, \hat{\bs \beta}_0,\hat{\bs \Psi}\big) > q_{1-\alpha}(\hat{\bs \beta}_0,\hat{\bs \Psi})$ where $q_{1-\alpha}(\hat{\bs \beta}_0,\hat{\bs \Psi})$ is the $1-\alpha$ quantile of $ T_{FAB}\big(\blt u, \hat{\bs \beta}_0,\hat{\bs \Psi}\big)$ where $\blt u \sim \text{Unif}(\mathbb S^{n_1-1})$. This quantile can be found by Monte Carlo simulation.
\end{enumerate}
This procedure results in a level-$\alpha$ test  since the independence of $\blt y_1$ and  $(\hat{\bs \beta}_0, \hat{\bs \Psi})$ implies that when $\bs \beta_1 = \blt 0$,
\begin{align*}
    P\big(  & T_{FAB}\big( \tfrac{\blt y_1}{ \Vert \blt y_1 \Vert}, \hat{\bs \beta}_0,\hat{\bs \Psi}\big)  > q_{1-\alpha}(\hat{\bs \beta}_0,\hat{\bs \Psi}) \big) 
    \\
    & = E\big(   P\big(  T_{FAB}\big( \tfrac{\blt y_1}{ \Vert \blt y_1 \Vert}, \bs \beta_0^*,\bs \Psi^*\big) > q_{1-\alpha}(\bs \beta_0^*,\bs \Psi^*) \big) | \hat{\bs \beta}_0 = \bs \beta_0^*, \hat{\bs \Psi} = \bs \Psi^*) \big)
    \\
    & = E(\alpha) = \alpha.
\end{align*}
We emphasize that this FAB test will be a level-$\alpha$ test, regardless of the validity of the linking model. All that is required for this test to have level-$\alpha$ is that $\blt y_1 /\Vert \blt y_1 \Vert$ must be independent of $(\hat{\bs \beta}_0, \hat{\bs \Psi})$ and  it must be uniformly distributed over the unit sphere.

Another desirable feature of the multigroup FAB test is that it approximately has the Bayes-optimal power if the linking model holds and if the the parameter estimates $(\hat{\bs \beta}_0,\hat{\bs \Psi})$ are close to $(\bs \beta_0,\bs \Psi)$. This follows by construction, since the likelihood ratio test that the multigroup FAB test approximates is the most powerful test marginally over the linking model. Conversely, mispecification of the linking model or poor estimates of $(\bs \beta_0,\bs \Psi)$ can result in a multigroup FAB test with sub-optimal power.

\subsection{Constructing a Data Dependent Prior Distribution}
\label{Sec:BetaPriorEst}
In this section we review some standard methods for obtaining estimates of $\bs \beta_0$ and $\bs \Psi$ from the random effects model
\begin{equation}
\label{eqn:REModel}
    \blt y_j | \bs \beta_j \sim N_{n_j}(\blt X_j \bs \beta_j,\sigma_j^2 \blt I), \;\; \bs \beta_j \sim N_p(\bs \beta_0,\bs \Psi), \;\;  j = 2,\ldots,m. 
\end{equation}
For simplicity, assume that $\sigma_2^2 = \cdots = \sigma_m^2$, where we denote the common value by $\sigma^2$. The marginal distribution of $\blt y_j$ is $N(\blt X_j \bs \beta_0,\bs \Sigma_j)$ with $\bs \Sigma_j = \blt X_j \bs \Psi \blt X_j^\top + \sigma^2 \blt I$. Under this marginal model for the $\blt y_j$'s, the maximum likelihood estimator of $\bs \beta_0$ is
\begin{equation}
\label{eqn:Beta0MLE}
        \hat{\bs \beta}_0 = \big( \sum_{j = 2}^m \blt X_i^\top  \hat{\bs \Sigma}_j^{-1} \blt X_j \big)^{-1}\big(\sum_{j = 2}^m \blt X_j^\top \hat{\bs \Sigma}_j^{-1} \blt y_j \big),
\end{equation}
where $\hat{\bs \Sigma}_j$ is the maximum likelihood estimator of $\bs \Sigma_j$.
The maximum likelihood estimators of $\sigma^2$ and $\bs \Psi$ have to be found numerically using, for example, the R package \texttt{lme4} \cite{Bates2015}. As a simpler alternative, moment based estimates of $\sigma^2$ and $\bs \Psi$ can be found which then can be substituted into the values of $\hat{\bs \Sigma}_j$ appearing in \eqref{eqn:Beta0MLE}. If $\blt P_j$ is the projection matrix onto $\text{col}(\blt X_j)$, the residual maximum likelihood estimate (REML) of $\sigma^2$ is given by
\begin{equation}\label{eqn:sigmaREML}
    \hat{\sigma}^2_{REML} = \big(\sum_{j  =2}^m (n_j - p)\big)^{-1} \sum_{j = 2}^m \Vert (\blt I - \blt P_j )\blt y_j \Vert^2 .
\end{equation}
A simple moment based estimate of $\bs \Psi$ is

\begin{align}
    \widehat{\bs \Psi} = \frac{1}{m-1} \sum_{j = 2}^m \bigg( &  (\blt X_j^\top \blt X_j)^{-1}\blt X_j^\top (\blt y_j - \blt X_j \hat{\bs\beta}_0)(\blt y_j - \blt X_j \hat{\bs \beta}_0)^\top \blt X_j (\blt X_j^\top \blt X_j)^{-1}  - 
    \nonumber
    \\
    & \hat{\sigma}^2_{REML}(\blt X_j^\top \blt X_j)^{-1}\bigg).
        \label{eqn:1stepPSi}
\end{align}
As $\widehat{\bs \Psi}$ depends on $\hat{\bs \beta}_0$ which in turn depends on $\widehat{\bs \Psi}$, an iterative procedure is needed to find suitable estimates. Such an iterative procedure can be initialized by taking $\widehat{\bs \Psi} = \blt I$ in \eqref{eqn:Beta0MLE}.

In summary, there is flexibility as to what estimates of $\bs \beta_0,\bs \Psi$ and $\sigma^2$ are used in the multigroup FAB procedure, as long as such estimates are independent of $\blt y_1$. If the values of the $n_j$'s and $p$ are large, the estimates in \eqref{eqn:sigmaREML} and \eqref{eqn:1stepPSi} may be easier to compute than the maximum likelihood estimates. The restriction that $\sigma_2^2 = \cdots = \sigma_m^2$ can also be lifted at the expense of additional computational effort. In this case either the MLE or direct analogues of the estimators in \eqref{eqn:sigmaREML} and \eqref{eqn:1stepPSi} could be used. Lifting this restriction may result in better estimates of the linking model parameters $\bs \beta_0$ and $\bs \Psi$ if the error variances across groups are different. However, the FAB test remains a level-$\alpha$ test regardless of the particular linking model parameter estimates chosen.

\subsection{Modelling the Error Variances}
As discussed in Section \ref{Sec:Theory}, the FAB test can approximately be viewed as a combination of the cone and $F$-tests. Roughly, if $\bs \beta_1$ is given the prior distribution $N(\bs \beta_0,\bs \Psi)$, the prior mean $\bs \beta_0$ determines the test direction of the cone test while the relative magnitudes of $\bs \Psi$ and $\sigma_0^2$ determine the how similar the FAB test is to either the $F$-test or the cone test. Recall that $\sigma_0^2$ was the location of the point mass prior distribution placed on $\sigma_1^2$. In this section we pursue a more sophisticated FAB test that adds to \eqref{eqn:REModel} the following linking model for the error variances $\sigma_j^2$:
\begin{equation}
\label{eqn:VarianceREModel}
    \sigma_1^2, \ldots, \sigma_m^2 \sim \text{i.i.d. } \text{Inverse-Gamma}(\alpha,\beta).
\end{equation}

Two additional steps are needed to incorporate this linking model into the multigroup FAB test previously discussed. First, the FAB test statistic \eqref{eqn:tfab} is altered to account for the new linking model \eqref{eqn:VarianceREModel}. Second, the observations in groups $2$ through $m$ are used to obtain estimates of $\alpha$ and $\beta$ to be used in this modified FAB test statistic. 

As before, by the Neyman-Pearson lemma, the FAB test statistic is the likelihood ratio test of the densities of $\blt y_1 /\Vert \blt y_1 \Vert$ under the new linking model and under the null hypothesis. Under the null hypothesis, $\blt y_1/ \Vert \blt y_1 \Vert$ remains uniformly distributed over the sphere, while under the linking model we have $\blt y_1/ \Vert \blt y_1 \Vert \, | \sigma_1^2 \sim AG(\bs \mu ,\bs \Sigma)$ with $\bs \mu = \blt X_1 \bs \beta_0$, $\bs \Sigma = \blt X_1 \bs \Psi \blt X_1^\top + \sigma_1^2 \blt I$ and $\sigma_1^2 \sim IG(\alpha,\beta)$. Therefore, the FAB test statistic is 
\begin{equation}
\label{eqn:IGTFAB}
    T_{IG-FAB}(\blt u) = \int_0^\infty \vert \bs \Sigma \vert^{-1/2}x^{-n}I_n(r)\exp((r^2 - \bs \mu^\top \bs \Sigma^{-1}\bs \mu)/2)\pi_{\alpha,\beta}(\sigma_1^2)d\sigma_1^2,  
\end{equation}
 where $x$ and $r$ are defined as in Section \ref{Sec:BasicFABTest} and $\pi_{\alpha,\beta}$ is the density of an $IG(\alpha,\beta)$ distribution. This statistic can be found via Monte Carlo approximation or numerical integration over the inverse gamma distribution. 
 
 Parameter estimates of $\bs \beta_0,\bs \Psi, \alpha$ and $\beta$ based on the data in groups $2$ through $m$ can be substituted into \eqref{eqn:IGTFAB}. One strategy for obtaining such estimates is to estimate $\bs \beta_0$ and $\bs \Psi$ via equations  \eqref{eqn:Beta0MLE} and \eqref{eqn:1stepPSi} as described in the previous section. Estimates of $\alpha$ and $\beta$ can be obtained by noting that $\Vert (\blt I - \blt P_j)\blt y_j \Vert^2 | \sigma_j^2 \sim \sigma_j^2 \chi^2_{n_j - p}$. To ease notation, define $e_j = \Vert (\blt I - \blt P_j)\blt y_j \Vert^2$ and $k_j = n_j - p$. For $\alpha > 2$, the  mean and variance of $e_j$, marginally over the distribution of $\sigma_j^2$, are given by 
 \begin{align*}
     E(e_j) &= \frac{k_j \beta}{\alpha - 1},\;\; E(e_j^2) = (2k_j + k_j^2) \frac{\beta^2}{(\alpha-1)(\alpha-2)}.
 \end{align*}
 If $e_{(1)} = \sum_{j = 2}^m e_j/(k_j(m-1))$ and $e_{(2)} = \sum_{j = 2}^m e_j^2/((2k_j + k_j^2)(m-1))$, then method of moments estimators for $\alpha$ and $\beta$ are found by solving the equations
 \begin{align*}
     e_{(1)} = \frac{\beta}{\alpha - 1}, \;\; e_{(2)} = \frac{\beta^2}{(\alpha-1)(\alpha-2)},
 \end{align*}
 yielding $\hat{\alpha} = (2e_{(2)} - e_{(1)}^2)/(e_{(2)} - e_{(1)}^2) > 2$ and $\hat{\beta} = e_{(1)}e_{(2)}/(e_{(2)} - e_{(1)}^2)$. These estimates are straightforward to compute. They tend to be more accurate when the $k_j$ are large as then $\chi^2_{k_j}/k_j \approx 1$ and thus $e_j/k_j \approx \sigma_j^2$. Other estimators for $\alpha$ and $\beta$ can also be used, however, they generally must be solved for numerically.
 
 The FAB test can have low power relative to the $F$-test if the estimate of $\bs \beta_0$ is poor. If this low power is a concern, a more conservative FAB test can be constructed by choosing estimates of $\alpha$ and $\beta$ so that $E(\sigma_1^2) = \beta/(\alpha-1)$ is large.

\subsection{Testing Other Linear Hypotheses}
\label{Sec:LinHypotheses}
In this section we describe how to test linear hypotheses in the multigroup regression model that are more general than the hypothesis $H:\bs \beta_1 = \blt 0$. For instance, in \eqref{eqn:wgm} it may be of interest to test if a subset of components of $\bs \beta_1$ are $0$, or to test the hypothesis $H:\bs \beta_1 = \bs \beta_2$ that the regression coefficients of two different groups are equal. 

Define $\bs \beta^\top_{1:l} = (\bs \beta_1^\top,\ldots,\bs \beta_l^\top)$ for $l < m$ and let $\blt A \in \mathbb R^{q \times lp}$, $\blt v \in \mathbb R^q$ with $\blt v \in \text{col}(\blt A)$. 
 A FAB procedure for testing the linear hypothesis $H:\blt A \bs \beta_{1:l} = \blt v$ on the regression coefficients of the first $l$ groups in the model \eqref{eqn:wgm} is described below.  We make the extra assumption in \eqref{eqn:wgm} that the error variances $\sigma_j^2$ are all equal to $\sigma^2$.   Define $\blt y_{j:k}^\top = (\blt y_j^\top,\ldots,\blt y_k^\top)$ and $\blt X_{j:k}^\top = [\blt X_j^\top,\ldots,\blt X_k^\top]$. The multigroup regression model \eqref{eqn:wgm} under the homoskedasticity assumption can be rewritten as
\begin{equation}
    \blt y_{1:l} \sim N(\blt X_{1:l} \bs \beta_{1:l},\sigma^2 \blt I), \;\; \blt y_j \sim N_{n_j}(\blt X_j \bs \beta_j,\sigma^2 \blt I) , \; j = l+1,\ldots,m.
\end{equation}
Let $S = \{\blt X_{1:l} \bs \beta_{1:l}: \blt A \bs \beta_{1:l} = \blt 0\}$ and take $\blt W$ to be a full-rank orthonormal matrix whose rows span the subspace $S^\perp$. Also take $\bs \beta^*_{1:l}$ to be a solution to the equation $\blt A \bs \beta_{1:l}^* = \blt v$ where we define $\bs \mu^* = \blt W \blt X_{1:l} \bs \beta_{1:l}^*$. By the definition of $\blt W$, the value of $\blt W \blt X_{1:l} \bs \beta_{1:l}^*$ is independent of the particular solution $\bs \beta_{1:l}^*$ chosen. Moreover, if the hypothesis $H$ is holds, this implies that the hypothesis $H^*:\blt W \blt X_{1:l}\bs \beta_{1:l} - \bs \mu^* = \blt 0$ also holds. When $\blt X_{1:l}$ is full-rank, or more generally when $\blt W \blt X_{1:l}$ is full-rank, these hypotheses are equivalent, meaning that $H^*$ is true if and only if $H$ is true.  

As  $\blt  W \blt y_{1:l} - \bs \mu^* \sim N(\blt W \blt X_{1:l} \bs \beta_{1:l} - \bs \mu^*,\sigma^2 \blt I)$, the hypothesis $H^*$ is identical to the hypothesis considered in Section \ref{Sec:BasicFABTest}, namely testing that the mean of the isotropic, multivariate normal random vector $\blt  W \blt y_{1:l} - \bs \mu^*$ is $\blt 0$. Applying the results from Section 2, if $\bs \beta_{1:l}$ is given the prior distribution  $\bs \beta_{1:l} \sim N_{lp}(  \blt 1 \otimes \bs \beta_{0},\blt I \otimes \bs \Psi)$, the FAB test of $H^*$ is exactly the likelihood ratio test with test statistic $\eqref{eqn:tfab}$ where $\blt u = (\blt W \blt y_{1:l} - \bs \mu^*)/\Vert  (\blt W \blt y_{1:l}  - \bs \mu^*)\Vert$ and $\bs \mu$ and $\bs \Sigma$ are the marginal mean and variance of $\blt W \blt y_{1:l} - \bs \mu^*$. We note that the common variance assumption is crucial in this setting to ensure that the distribution of $\blt u$ under the null hypothesis is pivotal. 

A data dependent prior distribution over $\bs \beta_{1:l}$ is obtained by finding estimates for $\bs \beta_{0}$ and $\bs \Psi$ in the linking model \eqref{eqn:agm}. Such estimates are found exactly as in Section \ref{Sec:BetaPriorEst}, using the observations $\blt y_{l+1},\ldots,\blt y_m$ that are not from the first $l$ groups. This procedure does not utilize any information from the portion $(\blt I - \blt W^\top  \blt W)\blt y_{1:l}$ of $\blt y_{1:l}$ that lies in $S$. If $\widetilde{\blt W}$ is a full-rank matrix with rows that span $S$, then marginally under the prior distribution \eqref{eqn:agm}, $ \widetilde{\blt W}\blt y_{1:l} \sim N(\widetilde{\blt W}\blt X_{1:l}( 1 \otimes \bs \beta_0),\sigma^2 \blt I + \widetilde{\blt W} (\blt I \otimes \bs \Psi) \widetilde{\blt W}^\top)$. The vector $\widetilde{\blt W}\blt y_{1:l}$ therefore does provide some useful information about the covariance structure of $\bs \beta_{1:l}$. This information is most easily incorporated into a maximum likelihood approach for estimating $\bs \beta_0$ and $\bs \Psi$. However, in hypotheses where $l \ll m$, such as the hypothesis that a subset of the components of $\bs \beta_1$ are $0$ when a large number of groups are present, we recommend using the simpler prior parameter estimates based on the observations $\blt y_{l+1},\ldots,\blt y_m$.     

\section{Example: Evaluating Standardized Test Scores}
\subsection{Overview of the ELS Data}
In this section we demonstrate the efficacy of the multigroup FAB test on educational outcome data. The 2002 educational longitudinal study (ELS) dataset includes demographic information of 15362 students from a collection of 751 schools across the United States in an effort to inform educational policy. We identify schools with ethnic disparities in educational outcomes by testing for mean differences in test scores by ethnicity after accounting for other variables. As some schools had only a small number of students who were surveyed, sharing information between the schools can help to improve the sensitivity of within-school testing procedures. On average, 20 students were surveyed per school, however 34 of the schools had less than 10 students who were surveyed. 

The response variable that we analyze is a (nationally) standardized composite math and reading score that is recorded for each student in the study. For each student we model the relationship between their test score and the following dependent variables: ethnicity, native language, sex, parental education and a composite index of the student's socio-economic status. Ethnicity is aggregated into four broad categories: Asian, Black, Hispanic and White. A separate linear regression model with the aforementioned predictor variables is used for each school. Independently across schools $j = 1,\ldots,751$, we take
\begin{equation}
\label{Eqn:FullELSModel}
    \blt y_j \sim N_{n_j}(\blt 
    Z_j \bs \alpha_j + \blt X_j \bs \beta_j , \sigma_j^2 \blt I), \;\; j = 1,\ldots, 751
\end{equation}
where $\blt y_j$ is the vector of test scores for the students in school $j$, $\bs \beta_{j}$ represents the regression coefficients for the ethnicity variables and $\bs \alpha_{j}$ represents the regression coefficients for the non-ethnicity variables in school $j$. For every school $j$, we test the hypotheses $H_j: \bs \beta_{j} = \blt 0$ by projecting out the non-ethnicity variables in \eqref{Eqn:FullELSModel}, a process which was described Section 4.4.

\begin{figure}
    \centering
    \includegraphics[height = 0.5\textwidth, width = \textwidth]{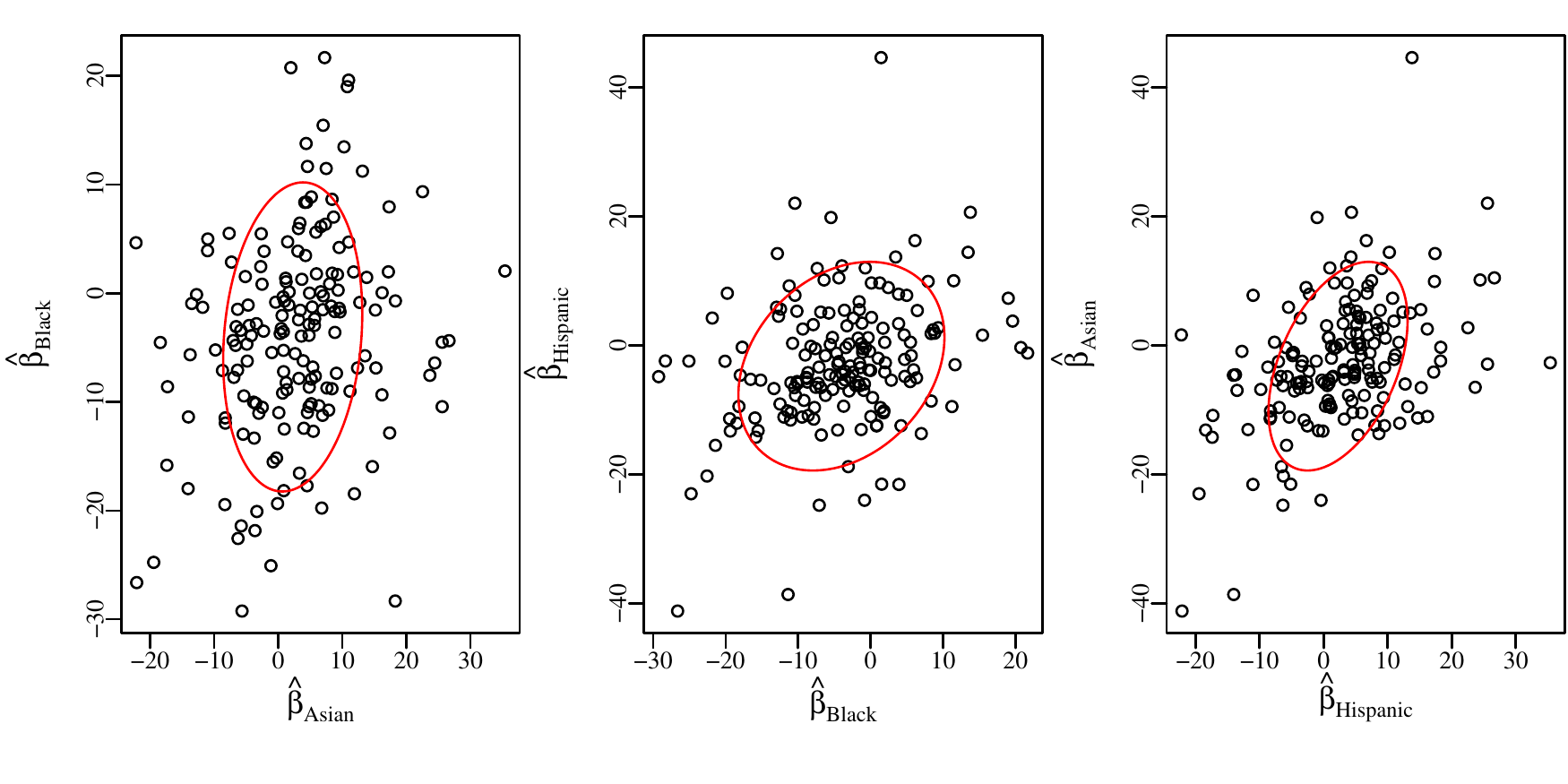}
    \caption{Pair plots of the least-squares estimates of the ethnicity regression coefficients along with $95\%$ confidence ellipses for $\bs \beta_{j}$.}
   \label{fig:CoefPairsPlot}
\end{figure}

Figure \ref{Fig:QQplots} in the Appendix illustrates normal QQ-plots of the residuals in the projected models for 9 different schools. These plots suggest that the projected regression model is a reasonable model for the ELS data. Figure \ref{fig:CoefPairsPlot} shows the distribution of the least square estimates of the $\bs \beta_{j}$ coefficients for schools with full-rank projected design matrices. From this figure, we conclude that it is not unrealistic to assume that the $\bs \beta_{j}$'s follow the multivariate normal linking model in \eqref{eqn:agm}. The $95\%$  confidence ellipses for $\bs \beta_{j}$ in Figure \ref{fig:CoefPairsPlot} are based on a method of moments estimate of $\bs \Psi$.

\begin{figure}
    \centering
    \includegraphics[width = \textwidth,height = 0.5\textwidth]{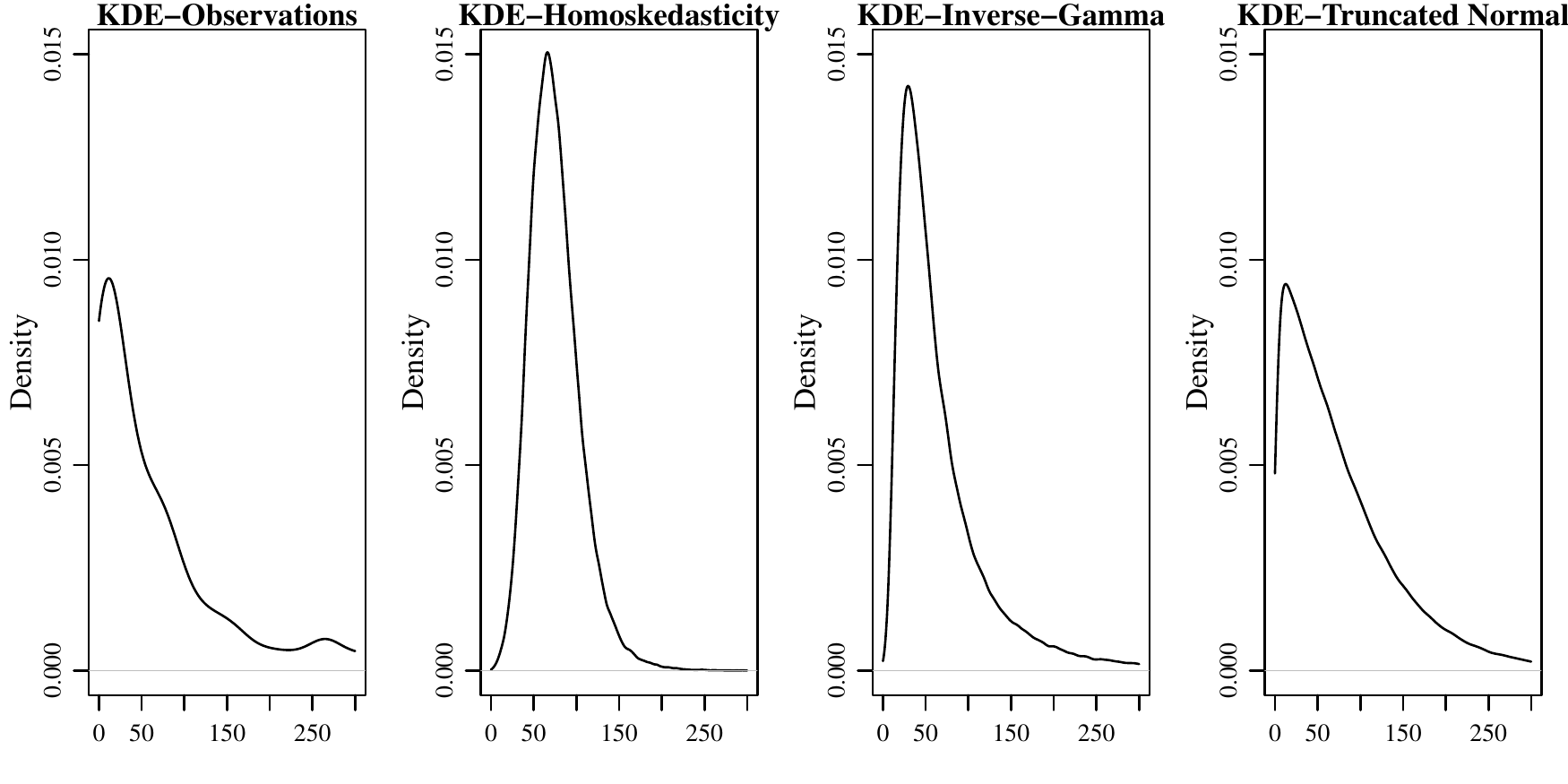}
    \caption{Kernel density estimates of the marginal distribution of the scaled squared-residuals under different linking models.}
    \label{fig:KernelDens}
\end{figure}

Lastly, we examine the empirical distribution of the scaled squared-residuals $e_j^2 = \Vert (\blt I - \blt P_j) \blt y_j \Vert^2$,  $j = 1,\ldots,751$ to determine a suitable linking model for the error variances.  A kernel density estimate of the marginal density of $e_j^2/(n_j - p), \; j = 1,\ldots,751$ is shown in Figure \ref{fig:KernelDens}. Under a linking model that assumes homoskedasticity, $\sigma^2 = \sigma_1^2 = \cdots = \sigma^2_{751}$, and $e_j^2/(n_j - p)$ has a chi-squared marginal distribution scaled by the constant $\sigma^2/(n_j - p)$. The second plot in Figure \ref{fig:KernelDens} sets $\hat{\sigma}^2 = \frac{1}{751}\sum_{j = 1}^{751}e_j^2/(n_j - p)$ and displays a kernel density estimate of the marginal distribution of $\{\hat{\sigma}^2w_1/(n_1 - p),\ldots,\hat{\sigma}^2w_{751}/(n_{751}-p)\}$ where this distribution is found by simulating $w_{j} \sim \chi^2_{n_j - p}$. It is apparent that the kernel density estimate of this marginal distribution does not match the kernel density estimate of the observed marginal distribution. Two other possible linking models are the inverse-gamma linking model $\sigma_j^2 \sim IG(\alpha,\beta)$ and the truncated normal linking model $\sigma_j^2 = \sigma_0^2 \vert z_j \vert, \; z_j \sim N(\mu,\tau^2)$. Fitting both of these models, it is seen in Figure 2 that the marginal density of $\sigma^2_jw_j/(n_j - p)$, $w_j\sim \chi^2_{n_j - p}, \;  j = 1,\ldots,751$ under the truncated normal linking model with $z_j \sim N(0.2,1.3)$ matches the observed marginal density more closely than the marginal density under the inverse-gamma linking model. 

\subsection{Methodology}
We test the hypothesis $H_j:\bs \beta_j  = \blt 0$ in the model \eqref{Eqn:FullELSModel} by projecting out the non-ethnicity variables as described in Section 4.4. That is, if $\blt W_j$ is a full-rank orthonormal matrix whose rows span $\text{col}(\blt Z_{j})^\perp$, the problem of testing $\bs \beta_{j} = \blt 0$  is reduced to testing this same hypothesis in the model $\blt W_j \blt y_j \sim N(\blt W_j \blt X_{j} \bs \beta_{j},\sigma_j^2 \blt I)$. The FAB or $F$-tests in the reduced model will have power greater than the level against some alternatives as long as $\blt W_j \blt X_{j} \neq  \blt 0$, a condition which holds for 634 of the 751 schools. Out of these 634 schools, 169 of them have full-rank projected design matrices $\blt W_j \blt X_{j}$. 
In the FAB tests we construct, the multivariate normal linking model \eqref{eqn:agm} is used to model $\bs \beta_{j}$.  Parameter estimates of $\bs \beta_0$ and $\bs \Psi$ are obtained using only the 169 schools that have the full-rank design matrices. The method of moments estimates of $(\bs \beta_0,\bs \Psi)$  described in Section \ref{Sec:BetaPriorEst} are used as estimates of these linking model parameters. As there are a sufficiently large number of observations from which to obtain parameter estimates, other estimation procedures will produce similar estimates.

 Two different linking model assumptions on the error variances $\sigma_j^2$ are used, resulting in two different FAB testing procedures. The first is simply the homoskedastic linking model that uses the FAB test statistic \eqref{eqn:tfab}, where the parameter $\sigma_0^2$ is estimated by the mean squared error of the reduced model, pooled across schools. The second truncated normal linking model assumption assumes that $\sigma^2_j = \sigma_0^2 \vert z_j \vert, \;\; z_j \sim N(0.2,1.3),  \;j = 1,\ldots,751$. We refer to these tests as FAB-HS and FAB-TN respectively. Again, the pooled mean squared error is used to estimate $\sigma_0^2$ in the truncated normal linking model. Denoting the marginal distribution of $\sigma_j^2$ by $\pi_{TN}(\sigma_j^2)$, analogous to equation \eqref{eqn:IGTFAB}, the FAB test statistic for school $j$ is given by the equation 
   \begin{equation}
\label{eqn:TNTFAB}
    T_{FAB-TN}(\blt u) = \int \vert  \hat{\bs \Sigma} \vert^{-1/2}x^{-n}I_n(r)\exp\big((r^2 -  \hat{\bs \mu}^\top  \hat{\bs \Sigma}^{-1}\hat{\bs \mu})/2\big)\pi_{TN}(\sigma_j^2)d\sigma_j^2,  
\end{equation}
where $\hat{\bs \Sigma} =  \blt W_j \blt X_{j}  \hat{\bs \Psi} \blt X_{j}^\top \blt W_j^\top + \sigma_j^2 \blt I$ and $x$ and $r$ are defined as in Section \ref{Sec:BasicFABTest} based on the data from school $j$. The integral in \eqref{eqn:TNTFAB} is approximated via Monte Carlo by drawing $100$ observations of $\sigma_j^2$ from the truncated normal distribution. 

To obtain the correct level when testing the hypothesis $H_j:\bs \beta_j = \blt 0$ for school $j$, theoretically, all of the estimates $( \hat{\bs \beta}_0, \hat{\bs \Psi},\hat{\sigma}_0^2)$ should be computed by leaving out the data from school $j$. For instance, when testing the hypothesis for school $2$, $\hat{\bs \beta}_0$ should be of the form $\hat{\bs\beta}_1 = \hat{\bs\beta}_1(\blt y_1,\blt y_3,\ldots\blt y_{751})$. However, as these parameter estimates are not significantly altered by leaving out school $j$, it is not necessary in practice to recompute these estimates for each school. Figure \ref{fig:PvalsRecomputed} in the Appendix displays the $p$-values for the homoskedastic FAB test where the linking model parameter estimates are either  recomputed when testing each hypothesis or the same parameter estimates are used for testing every hypothesis. It is seen that the $p$-values are nearly identical in these two cases, showing that at least when a large number of groups are present, it is not necessary to recompute the linking model parameter estimates for each hypothesis under consideration.

\subsection{Results}
\begin{figure}[ht]
    \centering
    \includegraphics[width = \textwidth, height = 0.5\textwidth]{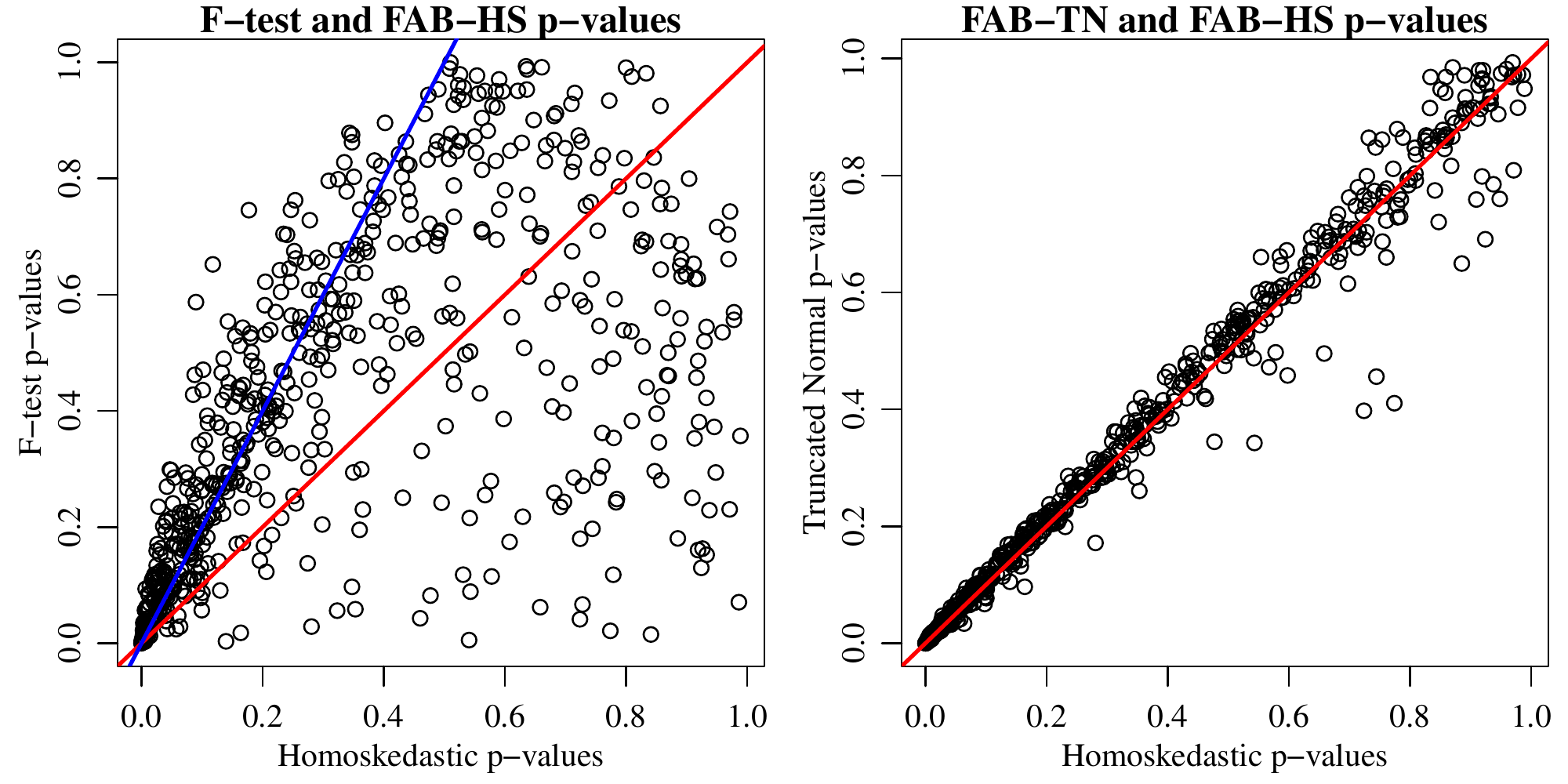}
    \caption{$p$-values for 634 schools under three different tests.}
    \label{fig:PvalScatter}
\end{figure}

We compare the empirical performance of three tests: the $F$-test, the homoskedastic FAB test and the truncated normal FAB test. In the second plot in Figure \ref{fig:PvalScatter} it is shown that the $p$-values for the homoskedastic and truncated normal FAB tests are nearly identical. This provides some evidence that the choice of the linking model over the $\sigma_j^2$'s is not of critical importance. In the first plot, when the $p$-values of the $F$-test are small, the $p$-values for the homoskedastic FAB test are typically less than the $F$-test $p$-values. These $p$-values are concentrated near the blue line with slope $2$ when the $p$-values of the FAB test are less than $0.5$.

Table \ref{Tab:Pvals} quantifies the extent to which the power of the FAB tests is greater than the power of the $F$-test. Across all of the 634 schools with non-zero projected design matrices, at a level of $0.05$, the FAB test rejects nearly twice as many hypotheses as the $F$-test. At a level of $0.01$, the relative improvement is even greater, with the FAB tests rejecting approximately $3$ times as many hypotheses. The average number of hypotheses rejected is higher on average for the 169 schools that have full-rank projected design matrices than for schools that do not. If $\blt W_j \blt X_{j}$ is not full-rank then the full vector $\bs \beta_j$ is not identifiable in \eqref{Eqn:FullELSModel}, rather only certain linear functions of $\bs \beta_j$ are identifiable. Therefore, if a school does not have a full-rank projected design matrix, not as much prior information can be leveraged in the FAB test. Effectively, prior information can only be used to directly inform the identifiable ``portion'' of $\bs \beta_j$.  For instance, if the first column of $\blt W_j \blt X_{j}$ happened to be $\blt 0$, the FAB test would not directly utilize the information from the other schools about the marginal distribution of the first component of $\bs \beta_{j}$.  

\begin{table}[ht]
\centering
\begin{tabular}{|r||rrr|}
  \hline
 & $F$-test & FAB-HS & FAB-TN \\ 
  \hline
  \hline
$\alpha = 0.05$, All Schools & 0.103 & 0.188 & 0.186 \\ 
  $\alpha = 0.01$, All Schools & 0.019 & 0.059 & 0.054 \\ 
  $\alpha = 0.05$, Full-rank Schools & 0.118 & 0.237 & 0.219 \\ 
  $\alpha = 0.01$, Full-rank Schools & 0.030 & 0.095 & 0.083 \\ 
   \hline
\end{tabular}
\caption{Percentage of hypotheses rejected at different levels, for schools with and without full-rank design matrices.}
\label{Tab:Pvals}
\end{table}

\begin{figure}[h]
    \centering
    \includegraphics[width = \textwidth, height = 0.5\textwidth]{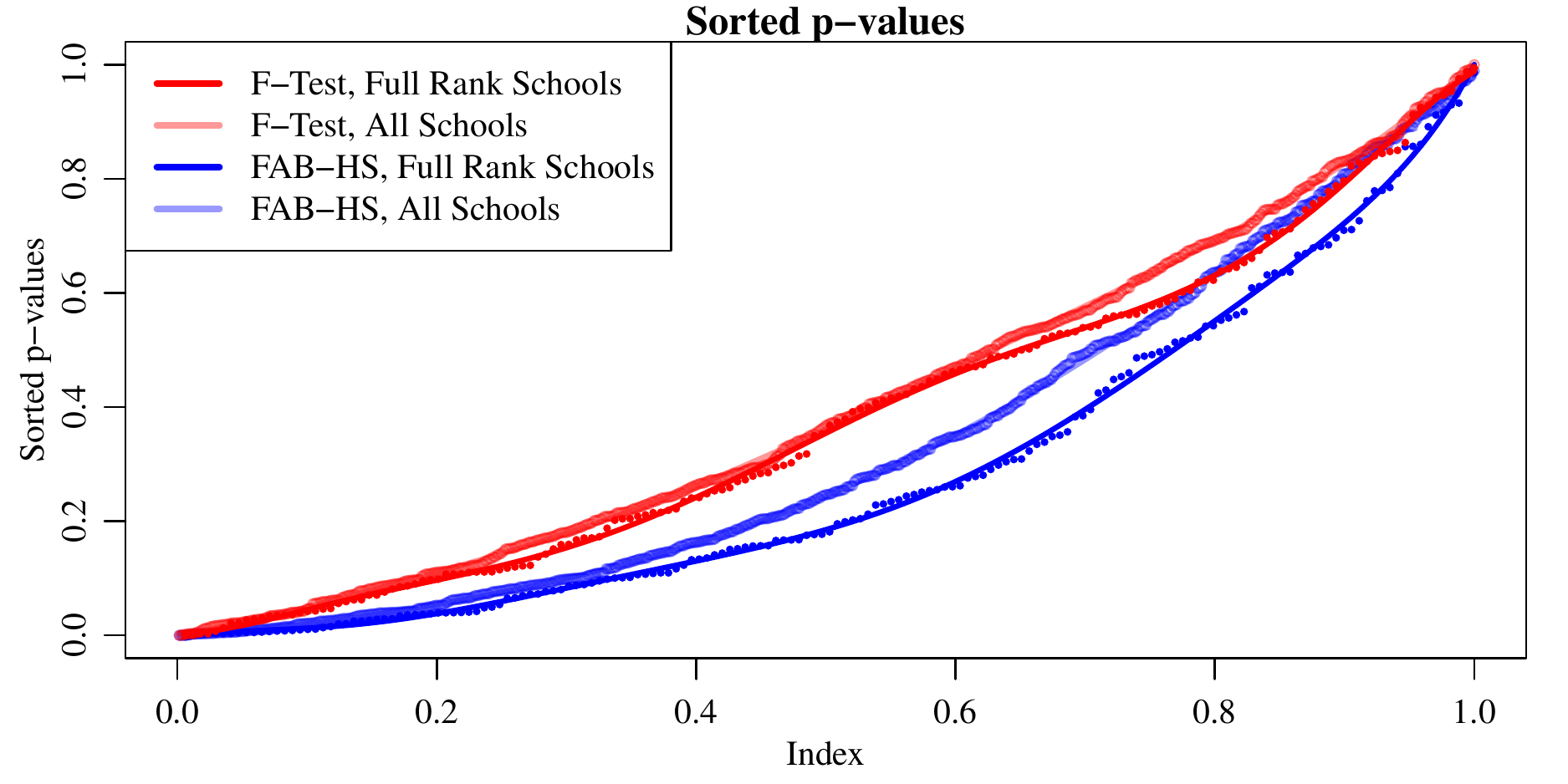}
    \caption{Sorted $p$-values for the $F$-test and FAB-HS Test}
    \label{fig:SortedPvalsFDR}
\end{figure}

Lastly, we examine the performance of the FAB tests relative to the $F$-test after controlling the false discovery rate (FDR). The Benjamini- Hochberg procedure controls the FDR at a level $\alpha$ by sorting the observed $p$-values $p_{(1)} < \cdots < p_{(m)}$ and rejecting all hypotheses $H_i$, $i \leq i^*$ where $i^*$ is the largest $i$ for which $p_{(i^*)} \leq \alpha i^*/m$ \cite{Benjamini1995}. This procedure is valid for $p$-values $\{p_1,\ldots,p_m\}$ that are independent, as in the $F$-test, or for $p$-values that satisfy certain types of positive dependence
\cite{Benjamini2001}. As the $p$-values obtained from the FAB test are constructed from estimates of the linking model parameters that are correlated, we expect the $p$-values to be positively dependent and the Benjamini-Hochberg procedure to be approximately valid for the FAB tests. Table \ref{Tab:BHFDR} in the Appendix compares the average number of hypotheses that are rejected for the $F$-test and FAB-HS test, controlling the FDR at various levels by the Benjamini-Hochberg procedure. The plot of the sorted $p$-values for the $F$-test and the FAB-HS test in Figure \ref{fig:SortedPvalsFDR} demonstrates that the empirical distribution of the FAB-HS $p$-values is stochastically smaller than the empirical distribution of the $F$-test $p$-values. In conclusion, whether controlling for size or for the FDR, the FAB tests in this setting are seen to be more powerful than the $F$-test.     

\section{Discussion}
In multigroup data analyses, inferences about one group can often  
be made more precise by utilizing data from the other groups. In this 
article, we have shown how a FAB test for a linear hypothesis 
involving one group may be constructed with the aid of data from the other 
groups, via a linking model that describes relationships among group-specific 
parameters. If the data from different groups are informative about 
each other's parameters, then such a test will have higher power than 
a direct test that does not make use of this information. Additionally, 
even if the linking model is incorrect or non-informative, the FAB test maintains 
exact type I error rate control. 

The FAB test statistic we have proposed is a function of the scale-invariant statistic $\blt y /\Vert \blt y \Vert$ which has a distribution that does not depend on $\sigma^2$. Effectively, by only considering scale-invariant test statistics, the null hypothesis has been reduced to a simple null hypothesis. This reduction to test statistics that are pivotal under the null hypothesis can be applied more broadly to test a wide variety of hypotheses. An interesting future direction is to develop FAB tests for nonparametric hypotheses, by restricting the test statistic to be a function of a pivotal quantity. One such example is to test the hypothesis that two distributions are equal using test statistics based on empirical distributions.

In nonparametric settings it may also be necessary to to consider more sophisticated prior distributions than the multivariate normal prior used in this article. In fact, the FAB prior distribution can itself be nonparametrically estimated. The possible utility of doing so is suggested by the ELS example where kernel density estimates of the scaled, squared-residuals are shown in Figure 2. Rather than using the parametric truncated normal or inverse-gamma linking models, a kernel density estimate of the density of the variance parameter could instead be used. However, if the model under consideration is parametric, for computational reasons it is preferable to keep the linking model as simple as possible.  

This article is focused entirely on testing the values of regression coefficients. In theory, the FAB test presented could be inverted to provide a confidence region for each vector of regression coefficients. Hoff and Yu \cite{Hoff2019} examine related confidence intervals for the elements of $\bs \beta$ in a regression model for a single group using shrinkage prior distributions. In the multigroup setting, properties such as the connectedness or convexity of such a confidence region found by inverting the FAB test warrant further study.

Another aspect of the multigroup FAB test that warrants further study are the multiple testing properties of this test. The multigroup FAB test controls the type I error rate for each group and thus controls the per-comparison error rate, but it does not control the family-wise error rate \cite{Dudoit2008}. Standard methods, such as the Bonferonni correction, can be used to control the family-wise error rate, although such methods may produce tests with low power. Similarly, it also is of interest to study methods for controlling the false discovery rate of the multigroup FAB test.

\newpage

\newpage

\section{Appendix}
\subsection{Proofs}

\begin{manualtheorem}{1}
This is a theorem.
Let $p_C(\blt y)$ be the $p$-value function for the cone test with rejection region \eqref{eqn:coneRejR} for testing the null hypothesis $H_0:\bs \beta =\blt 0$ where $\bs \mu \in \text{col}(\blt X)$ with $\Vert \bs \mu \Vert = 1$. If the observation $\blt y$ is of the form $\blt y = a\bs \mu + b \blt v$ with $\blt v \in \text{col}(\blt X)^\perp$, $a,b > 0$ then
\begin{align*}
    \frac{p_F(\blt y)}{p_C(\blt y)} = \frac{P\big( \sum_{i = 1}^p s_i^2 > c(\blt y) \big)}{P\big(s_1^2 > c(\blt y)\big)},
\end{align*}
where $(s_1^2,\ldots,s_n^2) \sim \text{Dirichlet}_n(\frac{1}{2},\ldots,\frac{1}{2})$ and $c(\blt y) = a^2/(a^2 + b^2)$.
In particular, for such a $\blt y$ the $p$-value ratio can be bounded below by
\begin{align*}
\frac{p_F(\blt y)}{p_C(\blt y)} \geq \frac{4}{(n-p)} \big(\frac{c(\blt y)}{1-c(\blt y)}\big)^{\frac{p-1}{2}} ,
\end{align*}
which tends to $\infty$ as $c(\blt y) \rightarrow 1$ if $1 < p < n$. 
\end{manualtheorem}
\begin{proof}
Without loss of generality the regression subspace $\text{col}(\blt X)$ can be assumed to be equal to $\text{span}\{\blt e_1,\ldots,\blt e_p\}$ where $\blt e_i$ is the $i$'th standard basis vector and $\bs \mu$ can be taken to be $\blt e_1$. Thus $\blt y = a \blt e_1 + b\blt v$ and $\blt v$ can be taken to be $\blt e_{p+1}$ since both the $F$ and cone test statistics only depend on $\blt y$ through the values of $a$ and $b \Vert\blt  v \Vert$. At $\blt y = a\blt e_1 + b\blt e_{p+1}$ the $F$ and cone test statistics are $(\Vert \blt P_X \blt y\Vert^2/p\big)/\big(\Vert (\blt I_n - \blt P_X) \blt y\Vert^2/(n-p)\big) = \big(a^2/p)/\big(b^2/(n-p)\big)$ and $\langle \blt y/\Vert \blt y \Vert,\blt e_1 \rangle = a/(a^2 + b^2)^{1/2}$ respectively. 

Under the null hypothesis $\blt y \sim N_n(\blt 0,\sigma^2\blt I_n)$ and the vector $\blt s = \blt y/\Vert \blt y\Vert$ has  $(s_1^2,\ldots,s_n^2) \sim \text{Dirichlet}_n(1/2,\ldots,1/2)$. The $F$-test $p$-value at $a\blt e_1 + b\blt e_{p+1}$ therefore has the form 
\begin{align*}
   p_F(\blt y) =  P\bigg(\frac{\sum_{i  =1}^p s_i^2}{\sum_{i = p+1}^ns_i^2} > \frac{a^2}{b^2}\bigg) = P\bigg(\sum_{i = 1}^ps_i^2 > \frac{a^2}{a^2 + b^2} \bigg),
\end{align*}
since $\Vert \blt P_X \blt Y \Vert^2/\Vert (\blt I_n -\blt P_X)Y\Vert^2 \overset{d}{=} (\sum_{i = 1}^ps_i^2)/(\sum_{i = p+1}^n s_i^2)$.
The conical test $p$-value is 
\begin{align*}
  p_C(\blt y) =   P\bigg(s_1 > \frac{a}{\sqrt{a^2 + b^2}}\bigg) = \frac{1}{2}P\bigg(s_1^2 > \frac{a^2}{a^2 + b^2}\bigg).
\end{align*}
We bound this ratio of beta probabilities below by
\begin{align*}
 \frac{p_F(\blt y)}{p_C(\blt y)} =    \frac{2P(\sum_{i = 1}^ps_i^2 > c)}{P(s_1^2 > c)} & = \frac{2\Gamma(\frac{1}{2})\Gamma(\frac{(n-1)}{2}) }{\Gamma(\frac{p}{2})\Gamma(\frac{(n-p)}{2})} \frac{\int_c^1 x^{\frac{p}{2}-1}(1-x)^{\frac{n-p}{2}-1}dx  }{\int_c^1 x^{\frac{1}{2}-1}(1-x)^{\frac{n-1}{2} - 1}dx} 
    \\
    & \geq \frac{2 c^{\frac{p}{2}-1}\int_c^1 (1-x)^{\frac{n-p}{2}-1}dx  }{c^{-\frac{1}{2}}(1-c)^{\frac{n-1}{2}}} = \frac{4}{(n-p)} c^{\frac{p-1}{2}}(1-c)^{\frac{1-p}{2}}.
\end{align*}
\end{proof}

\begin{manualtheorem}{2}
Consider the sequence of models $\mathcal{M}_n:\blt y_n \sim N_n(\blt X_n \bs \beta_n,\sigma^2\blt I)$ where $\bs \beta_n \in \mathbb{R}^{p_n}$. Define $\rho_n(c_n,p_n,\sigma^2)$ to be the power of the level $\alpha$ $F$-test of the null hypothesis $H_{0,n}:\bs \beta_n = \blt 0$ under the alternative hypothesis that has $\Vert \blt X_n \bs \beta_n\Vert = c_n$. If $p_n = p_0$ and $c_n = c_0$ are constants then $\alpha < \liminf_{n \rightarrow \infty} \rho_n(c_0,p_0,\sigma^2) < 1$. If $\gamma \in (0,1)$ then  $\liminf_{n \rightarrow \infty} \rho_n(c_0,\lfloor \gamma n\rfloor,\sigma^2)  = \alpha$ and if $c_n = n^{1/4}$ then the $F$-test has limiting power $\liminf_{n \rightarrow \infty} \rho_n(n^{1/4},\lfloor \gamma n \rfloor,\sigma^2) \in (\alpha,1)$.     
\end{manualtheorem}
\begin{proof}
The $F$ statistic is
\begin{align*}
    F_n(\blt y_n) =  \frac{\Vert \blt P_{X_n}\blt y_n\Vert^2/p_n }{\Vert (\blt I_n - \blt P_{X_n})\blt y_n\Vert^2/(n - p_n)}
\end{align*}
In the first regime where $p_n = p_0$ and $c_n = c_0$ this statistic converges in probability to a $\chi^2_p$ distribution under the null hypothesis and a $\chi^2(c_0^2)$ distribution under the alternative hypothesis since the denominator converges in probability to $\sigma^2$ and the numerator follows a chi-squared distribution. If $Z \sim \chi^2_p(c_0^2)$ then $\liminf_{n \rightarrow \infty}\rho_n(c_0,p_0,\sigma^2) = P(Z > \chi^2_{p_0,1- \alpha}) \in (\alpha,1)$ as needed.

Next, consider the case where $p_n = \lfloor \gamma n \rfloor$. Under the alternative hypothesis the numerator of the $F$-statistic has a $\sigma^2 \chi^2_{p_n}(c_n^2/\sigma^2)$ distribution with the stochastic representation $\sigma^2 (z_1 + c_n/\sigma)^2 + \sigma^2 \sum_{i = 2}^{p_n}z_i^2 = c_n^2 + 2z_1c_n\sigma + \sigma^2 \sum_{i = 1}^{p_n}z_i^2$ where $z_i \overset{i.i.d.}{\sim} N(0,1)$. Define $w_n = (n - p_n)\sigma^2 \sum_{i = 1}^{p_m}(z_i^2 - 1)/(\sqrt{2p_n}\Vert (\blt I_n -\blt P_{X_n})\blt y_n\Vert^2)$. As $n \rightarrow \infty$, $w_n \overset{d}{\rightarrow} N(0,1)$ and consequently, $\sqrt{p_n/2}(F_{p_n,n- p_n,1-\alpha} - 1) \rightarrow z_{1 - \alpha}$ where $z_{1 - \alpha}$ is the $1-\alpha$ standard normal quantile. Consequently, 
\begin{align*}
    \liminf_{n \rightarrow \infty} \; &  \rho_n(c_n,p_n,\sigma^2) 
    \\
    & =  \liminf_{n \rightarrow \infty}\; P\bigg( w_n +  \frac{ c_n^2/\sqrt{p_n} + 2z_1c_n \sigma/\sqrt{p_n}}{\sqrt{2}\Vert (\blt I_n - \blt P_{X_n})\blt y_n\Vert^2/(n-p_n)}  >
     \sqrt{p_n/2}(F_{p_n,n - p_n,1-\alpha} - 1) \bigg)
\end{align*}
If $c_n = o(p_n^{1/4})$ then $ \liminf_{n \rightarrow \infty} \rho_n(c_n,p_n,\sigma^2) = P(N(0,1) > z_{1-\alpha}) =  \alpha$, while if $c_n = cp_n^{1/4}$ then $ \liminf_{n \rightarrow \infty} \rho_n(c_n,p_n,\sigma^2) = P\big(N(0,1) > z_{1-\alpha} - c^2/(\sqrt{2} \sigma^2)\big) \in (\alpha,1)$ and the claims follow.
\end{proof}

\begin{manualtheorem}{3}
Consider the sequence of models $\mathcal{M}_n: \blt y_n \sim N_n(\blt X_n \bs \beta_n,\sigma^2\blt I)$ where $\bs \beta_n \in \mathbb{R}^{p_n}$. Define $\rho_n(\bs \mu_n,\blt v_n,p_n,\sigma^2)$ to be the power of the level-$\alpha < 1/2$  cone test of the null hypothesis $H_{0,n}:\bs \beta_n = \blt 0$ with rejection region $\{\blt y:\langle \blt y/\Vert \blt y\Vert,\bs \mu_n\rangle > q_{n,1-\alpha}\}$ where $q_{n,1-\alpha}$ is an appropriate level-$\alpha$ quantile, $\bs \mu_n \in \mathbb{S}^{n-1}$ and $\blt v_n = \blt X_n \bs \beta_n$ under the alternative hypothesis. If $\Vert \blt v_n\Vert^2  = c_0$ is constant and the mean direction of the cone test is nearly correctly specified so that $\Vert \bs \mu_n - \blt v_n/c_0 \Vert = o(n^{-1/2})$ then $\liminf_{n \rightarrow \infty} \rho_n(\bs \mu_n,\blt v_n,p_n,\sigma^2) \in (\alpha,1)$ where the power function does not depend on $p_n$. If $\Vert \blt v_n\Vert = n^{1/4}$ and if $\Vert \bs \mu_n - (\blt v_n/\Vert \blt v_n \Vert) \Vert = (n^{-1/4} - an^{-1/2}) = O(n^{-1/4})$ for some $a > 0$ then  $\liminf_{n \rightarrow \infty} \rho_n(\bs \mu_n,\blt v_n,p_n,\sigma^2) \in (\alpha,1)$ and if $\Vert \bs \mu_n - (\blt v_n/\Vert \blt v_n \Vert) \Vert = o(n^{-1/4})$ then  $\liminf_{n \rightarrow \infty} \rho_n(\bs \mu_n,\blt v_n,p_n,\sigma^2) = 1$.
\end{manualtheorem}
\begin{proof}
Define $c_n = \Vert \blt X_n \bs \beta_n\Vert$ and without loss of generality we assume that $\blt X_n \bs \beta_n = c_n \blt e_1 \in \mathbb{R}^n$, where $\blt e_1$ is the first standard basis vector.  
Under the alternative hypothesis the conical test statistic has the stochastic representation 
\begin{align*}
    \langle \frac{\blt y_n}{\Vert \blt y_n\Vert},\bs \mu_n\rangle & = \langle \frac{\blt z_n + c_n \blt e_1}{\sqrt{\Vert \blt z_n \Vert^2 + 2c_n z_{n,1} + c_n^2}},\blt e_1 + (\bs \mu_n - \blt e_1) \rangle
    \\
    & \geq \frac{z_{n,1} + c_n}{\sqrt{\Vert \blt z_n \Vert^2 + 2c_n z_{n,1} + c_n^2}} - \Vert \bs \mu_n - \blt e_1 \Vert,
\end{align*}
where $\blt z_n \sim N_n(\blt 0,\sigma^2\blt I_n)$, and $z_{n,i}$ is the $i$'th component of $\blt z_n$. Thus 
\begin{align*}
    \rho_n(\bs \mu_n,c_n\blt e_1, p_n,\sigma^2) \geq P\bigg(\frac{z_{n,1} + c_n}{\sqrt{\Vert \blt z_n \Vert^2 + 2c_nz_{n,1} + c_n^2}} > q_{n,1-\alpha} + \Vert \bs \mu_n - e_1 \Vert\bigg).
\end{align*}
By the law of large numbers $\sqrt{n}z_{n,1}/\Vert \blt z_n \Vert \overset{d}{\rightarrow} N(0,1)$ and thus $\sqrt{n}q_{n,1-\alpha} \rightarrow z_{1-\alpha}$ where $z_{1-\alpha}$ is a standard normal quantile. Then assuming that $\Vert\bs \mu_n -\blt e_1 \Vert = o(n^{-1/2})$ and $c_n = c_0$ is constant
\begin{align*}
     \liminf_{n \rightarrow \infty}& \; \rho_n(\bs \mu_n,c_0\blt e_1,p_n,\sigma^2)&
     \\
    & \geq \liminf_{n \rightarrow \infty} \; P\bigg(\frac{z_{n,1} + c_0}{\sqrt{\frac{1}{n}\Vert \blt z_n \Vert^2 + \frac{2c_0}{n} z_{n,1} + \frac{c_0^2}{n} }} > \sqrt{n}q_{n,1-\alpha} + \sqrt{n}\Vert \bs \mu_n - e_1 \Vert\bigg)
     \\
     & =  P\bigg( z_{n,1} + c_0 > z_{1 - \alpha}  \bigg) > \alpha
\end{align*}
Next assume that $c_n = n^{1/4}$ and $\Vert \bs \mu_n -\blt e_1 \Vert = (n^{-1/4} - an^{-1/2})$ so that
\begin{align*}
       \liminf_{n \rightarrow \infty}\; & \rho_n(\bs \mu_n,c_n \blt e_1,p_n,\sigma^2) 
       \\
       & \geq \liminf_{n \rightarrow \infty}P\bigg(\frac{z_{n,1} + c_n}{\sqrt{\frac{1}{n}\Vert \blt z_n \Vert^2 + \frac{2}{n^{3/4}} z_{n,1} + \frac{1}{\sqrt{n}} }} > \sqrt{n}q_{n,1-\alpha} + \sqrt{n}\Vert \bs \mu_n - \blt e_1 \Vert\bigg)
     \\
     & = \liminf_{n \rightarrow \infty} P\bigg( z_{n,1}  > z_{1 - \alpha} + n^{1/4}(n^{1/4}\Vert \blt v_n - \blt e_1 \Vert - 1) \bigg)  
     \\
     & =  P\bigg( z_{n,1}  > z_{1 - \alpha} - a\bigg) > \alpha
\end{align*}
The result for $\Vert \bs \mu_n  - \blt e_1 \Vert = o(n^{-1/4})$ immediately follows as well since $\lim_{n \rightarrow \infty}\sqrt{n}\Vert \bs \mu_n - \blt e_1\Vert - n^{1/4} = -\infty$.
\end{proof}

\begin{manualtheorem}{4}
Under the prior distribution $\bs \beta \sim N_p(\bs \beta_0,\gamma (\blt X^\top \blt X)^{-1})$, the FAB test is asymptotically equivalent to the F-test as $\gamma \rightarrow \infty$. That is, the probability that the F and FAB tests lead to the same conclusion under the null hypothesis or any alternative hypothesis tends to $1$ as $\gamma \rightarrow \infty$.
\end{manualtheorem}
\begin{proof}
Let $R_{\alpha,F} = \{\blt y: F(\blt y/\Vert \blt y \Vert) > c_{\alpha,F}\}$ and $R_{\alpha,\gamma} =  \{\blt y: T_{FAB,\gamma}(\blt y/\Vert \blt y \Vert) > c_{\alpha,\gamma}\}$ be the rejection regions for the F and FAB tests respectively $(\alpha < 0.5$). It suffices to show that $\bigcap_{k > 0 }\bigcup_{\gamma > k}  R_{\alpha,F} \Delta R_{\alpha,\gamma}$ has Lebesgue measure $0$, where $A \Delta B$ is the symmetric difference of sets.

As $\gamma \rightarrow \infty$, if $\Vert (\blt I - \blt P) \blt u \Vert^2 > 0$, then
\begin{equation}
    T_{FAB,\gamma}(\blt u) \rightarrow n\log(\sigma_0) -\frac{n}{2}\log( \Vert (\blt I - \blt P)\blt u \Vert^2) + \log\big(\int_0^\infty z^{n-1}\exp(-z^2)dz\big),
    \label{eqn:TFabGammaLimit}
\end{equation}
where an appeal to the dominated convergence theorem is used to take the limit $w \rightarrow 1$ inside the integral in \eqref{eqn:FABStatSimplified}. Call the statistic on the right hand side of \eqref{eqn:TFabGammaLimit}  $T_{FAB,\infty}(\blt u)$. If $\blt z \sim \text{Uniform}(\mathbb S^{n-1})$ then $T_{FAB,\gamma_n}(\blt z) \overset{a.s.}{\rightarrow} T_{FAB,\infty}(\blt z)$ for any sequence $\gamma_n \rightarrow \infty$ since $P(\Vert (\blt I - \blt P)\blt z \Vert^2 = 0) = 0$. If $c_{\alpha,\infty}$ is the $(1-\alpha)$-quantile of $T_{FAB,\infty}(\blt z)$ then it is claimed that for any $\gamma_n \rightarrow \infty$, $c_{\alpha,\gamma_n} \rightarrow c_{\alpha,\infty}$. If this were not the case then there would exist a sequence $\gamma_n \rightarrow \infty$ with $\vert c_{\alpha,\gamma_n} - c_{\alpha,\infty} \vert > \epsilon$ for all $n$. However, this is not possible since $T_{FAB,\gamma_n}(\blt z) \overset{P}{\rightarrow} T_{FAB,\infty}(\blt z)$ and $T_{FAB,\infty}(\blt z)$ has a continuous distribution with positive density on the interior of its support (implying that its quantiles are unique).

 We have shown above that if $\gamma_n \rightarrow \infty$ then $\lim_n T_{FAB,\gamma_n}(\blt u) - c_{\alpha,\gamma_n} = T_{FAB,\infty}(\blt u) - c_{FAB,\infty}$ for all $\blt u$ with $\Vert (\blt I - \blt P)\blt u \Vert^2 > 0$. As $T_{FAB,\infty}(\blt u) - c_{\alpha,\infty} > 0$ if and only if $F(\blt u) - c_{\alpha,F} > 0$, all $\blt y \in \bigcap_{k > 0 }\bigcup_{\gamma > k}  R_{\alpha,F} \Delta R_{\alpha,\gamma}$ must satisfy $\Vert (\blt I - \blt P) \blt y  \Vert^2 = 0$. As this set has Lebesgue measure $0$ (assuming that $\blt P \neq \blt I$) this completes the proof. 
  
\end{proof}

\subsection{Tables and Figures}
\begin{figure}[h]
    \centering
\includegraphics[width = \textwidth, height = .6\textwidth]{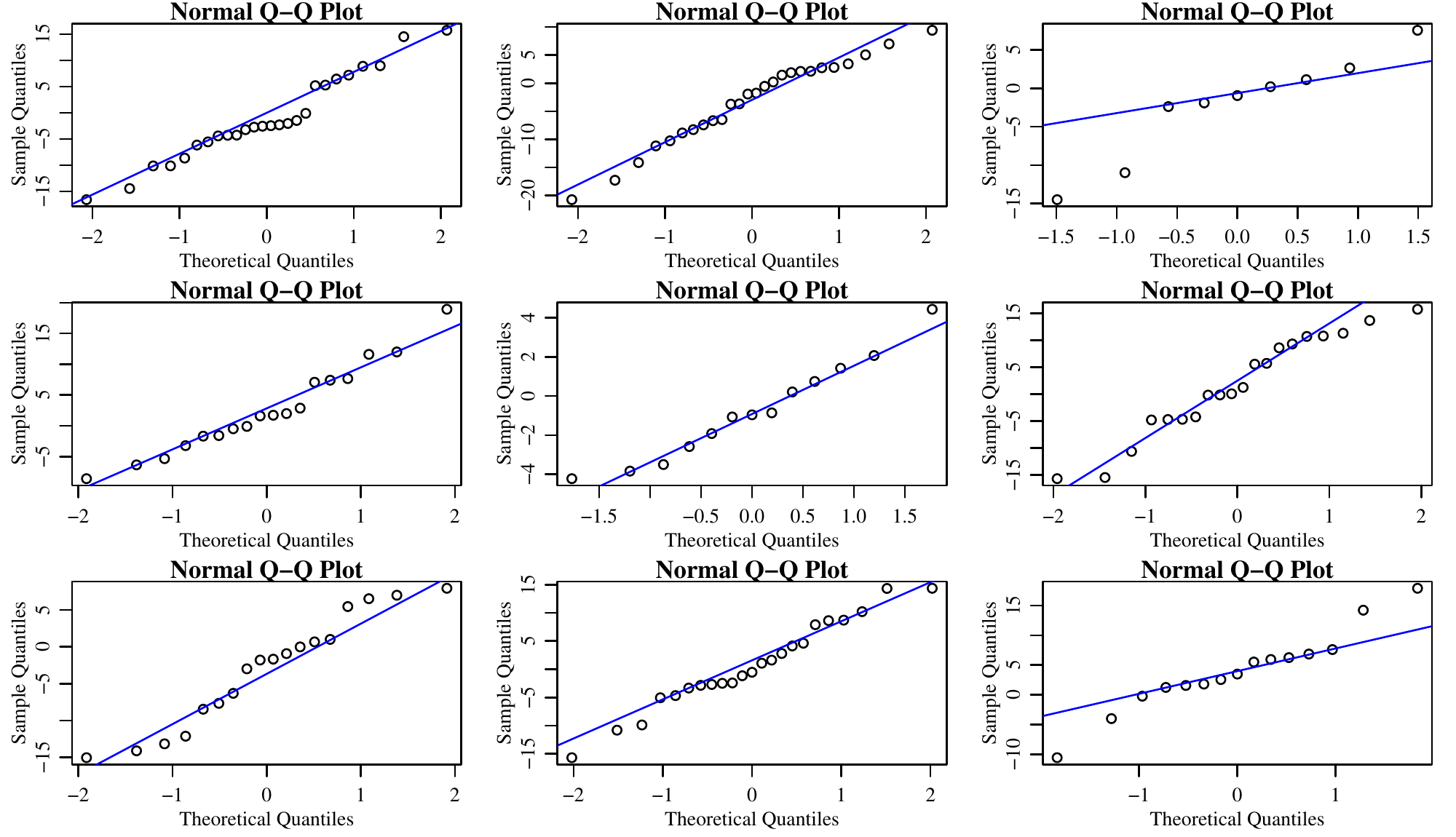}
    \caption{Normal Q-Q plots for $9$ different schools}
    \label{Fig:QQplots}
\end{figure}

\begin{figure}[h]
    \centering
    \includegraphics[width = 0.65\textwidth, height = 0.455\textwidth]{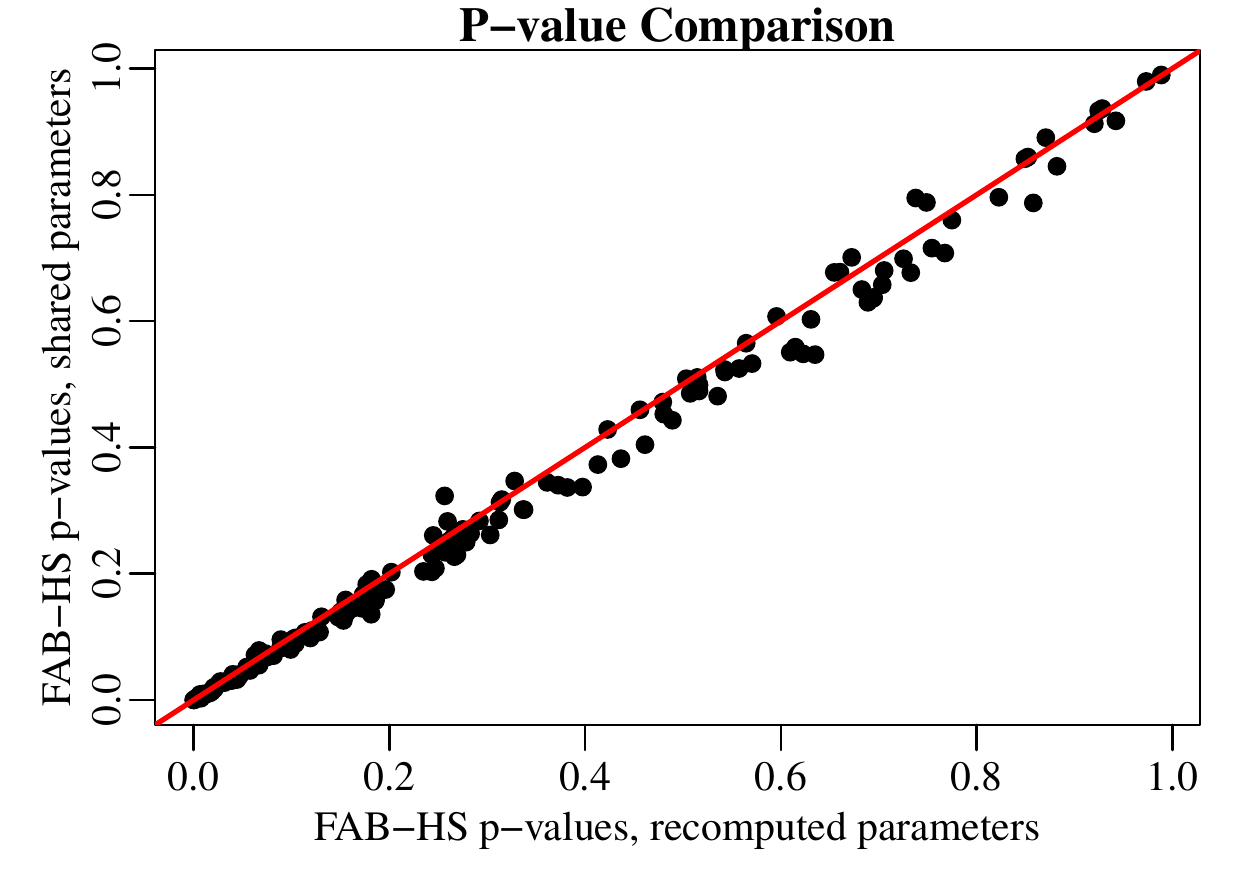}
    \caption{Comparison of the FAB-HS $p$-values when the prior parameter estimates of $\bs \beta_0,\bs \Psi,\sigma_0^2$ are either recomputed for each school or common estimates are used across all schools. }
    \label{fig:PvalsRecomputed}
\end{figure}

\begin{table}[h]
\centering
\begin{tabular}{|r||rrrr|}
  \hline
 & FAB-HS, All & FA-HS, full-rank & $F$-test, All & $F$-test, full-rank\\ 
  \hline
  \hline
$\alpha = 0.1$ & 0.01 & 0.11 & 0.00 & 0.01 \\ 
  $\alpha = 0.2$ & 0.09 & 0.24 & 0.00 & 0.02 \\ 
  $\alpha  = 0.5$ & 0.52 & 0.68 & 0.12 & 0.28 \\ 
   \hline
\end{tabular}
\caption{Proportion of hypotheses rejected by the Benjamini-Hochberg Procedure across either all schools or the schools with full-rank design matrices.}
\label{Tab:BHFDR}
\end{table}

\end{document}